\documentclass[prd,floatfix,nofootinbib,eqsecnum,showpacs]{revtex4}
\usepackage{epsfig}
\usepackage{amssymb,amsmath}
\newcommand{\eq}[1]{Eq.~(\ref{#1})}

\newcommand{\be}{\begin{equation}}
\newcommand{\ee}{\end{equation}}
\newcommand{\bea}{\begin{eqnarray}}
\newcommand{\eea}{\end{eqnarray}}
\newcommand{\ben}{\begin{eqnarray*}}
\newcommand{\een}{\end{eqnarray*}}

\newcommand{\DS}{Dyson-Schwinger }
\newcommand{\BS}{Bethe-Salpeter }

\newcommand{\w}{\omega}
\newcommand{\e}{\varepsilon}
\newcommand{\al}{\alpha}
\newcommand{\ba}{\beta}
\newcommand{\ga}{\gamma}
\newcommand{\G}{\Gamma}
\newcommand{\de}{\delta}

\newcommand{\si}{\sigma}

\newcommand{\ro}{\rho}
\newcommand{\la}{\lambda}
\newcommand{\La}{\Lambda}

\newcommand{\ta}{\tau}

\newcommand{\pd}{\partial}

\newcommand{\s}[2]{{#1}\!\cdot\!{#2}}
\newcommand{\ov}[1]{\overline{#1}}
\newcommand{\dk}[1]{\,\,\,\raisebox{-0.4ex}{\large $\bar{}$}\!\!d\,{#1}\,}

\newcommand{\ev}[1]{<\!\!{#1}\!\!>}

\begin{document}
\title{Bethe-Salpeter equation at leading order in Coulomb gauge}
\author{P.~Watson}
\author{H.~Reinhardt}
\affiliation{Institut f\"ur Theoretische Physik, 
Universit\"at T\"ubingen, Auf der Morgenstelle 14, 
D-72076 T\"ubingen, Deutschland}
\begin{abstract}
The Bethe-Salpeter equation and leptonic decay constants for pseudoscalar 
and vector quark-antiquark mesons with arbitrary quark masses are studied 
in Coulomb gauge, under a leading order truncation.  As input, we use a 
pure linear rising potential, supplemented by a contact term arising from 
the conservation of total color charge.  It is shown how the equations can 
be written in terms of manifestly finite functions, despite the infrared 
singular interaction.  The resulting equations are solved numerically.  
Both the pattern of dynamical chiral symmetry breaking and the leading 
order heavy quark limit are visible.
\end{abstract}
\maketitle
\section{Introduction}
In the theory of strong interactions, quantum chromodynamics (QCD), there 
are two outstanding issues of importance: confinement and dynamical chiral 
symmetry breaking.  Both of these phenomena are highly nontrivial and are 
reflected in the nonperturbative regime of the theory.  Perhaps the most 
naive picture of confinement, at least applied to the quark sector of QCD, 
is the idea that there exists a linear rising potential between quarks at 
large separation.  If a quark is pulled away from its parent hadron, the 
energy of the system rises until it is energetically favorable that a 
quark-antiquark pair is created and forming a new system of two hadrons 
whose constituents remain hidden from view in the same way as for the 
original.  In momentum space, such a linear rising potential corresponds 
to a strongly infrared enhanced interaction.  In Coulomb gauge, lattice 
results \cite{Iritani:2010mu,Voigt:2008rr,Nakagawa:2011ar,Nakagawa:2009is,
Quandt:2008zj,Cucchieri:2007uj,Langfeld:2004qs,Cucchieri:2000gu} 
show that the temporal component of the nonperturbative gluon propagator 
(that mediates the quark-antiquark interaction) indeed exhibits such an 
infrared enhancement.

The inclusion of an infrared singularity, such as that of the 
aforementioned temporal gluon propagator, into calculations of finite, 
physical observables is not trivial.  Because the singularity occurs for 
vanishing momentum, a natural place to investigate would be a 
nonperturbative, continuum formalism, such as that provided by the 
\DS and \BS equations\footnote{An alternative approach is to consider 
the Coulomb gauge Hamiltonian: see for example, 
Refs.~\cite{LlanesEstrada:2001kr,LlanesEstrada:2004wr,Schutte:1985sd,
Szczepaniak:2001rg,Szczepaniak:2003ve,Feuchter:2004mk,
Reinhardt:2004mm,Fontoura:2012mz} and references therein.}.  The Coulomb 
gauge \DS equations for QCD have been derived in both the first and 
second order formalisms 
(see Refs.~\cite{Watson:2006yq,Watson:2007vc,Popovici:2008ty} and 
references therein) and various aspects explored, such as the one-loop 
perturbative behavior \cite{Watson:2007mz,Watson:2007vc,Popovici:2008ty}, 
the Slavnov-Taylor identities \cite{Watson:2008fb} and the emergence of a 
nonlocal constraint on the total color charge \cite{Reinhardt:2008pr}.  
Additionally, the case of heavy quarks in the rest frame has also been 
studied \cite{Popovici:2010mb,Popovici:2010ph,Popovici:2011yz}.  These 
latter studies show that in the absence of pure Yang-Mills corrections, 
the Coulomb gauge rest frame heavy quark limit (where pair creation is 
absent, such that the linear rising potential remains valid at very large 
separations) is described precisely by an interaction consisting of single 
exchange of a temporal gluon.  The infrared singularity associated with 
this interaction is cancelled only when considering color singlet 
(meson or baryon) bound states, otherwise the objects 
(propagators or color nonsinglet states) are unphysical.

The issue of dynamical chiral symmetry breaking was studied in Coulomb 
gauge many years ago (see, for example 
Refs.~\cite{Govaerts:1983ft,Adler:1984ri,Alkofer:1988tc}).  It was found 
that with an infrared enhanced temporal gluon interaction, chiral symmetry 
breaking does occur, although the chiral condensate and pion leptonic 
decay constant are too small (it has been demonstrated that the coupling 
of the quarks to transverse spatial gluons results in a significantly 
larger condensate \cite{Pak:2011wu}).  Recently, it has been shown that 
the twin limits of the chiral and heavy quark propagator in Coulomb gauge, 
with an infrared enhanced interaction, can be accommodated within a 
single leading order truncation scheme for the \DS equations 
\cite{Watson:2011kv}\footnote{The interplay of the chiral and heavy quark 
symmetries has also been studied in an extended Nambu--Jona-Lasino model 
\cite{Ebert:1994tv}.}.  The truncation scheme is centered around an 
Ansatz for the Coulomb kernel occurring in the action: this term originates 
from the resolution of Gauss' law and is responsible for the temporal 
component of the interaction between color charges, i.e., it is intimately 
connected to the temporal component of the gluon propagator.  The 
interaction between quarks and the spatial components of the gluon field, 
the three-, and the four-gluon vertices are all neglected under this 
leading order truncation scheme.

In this paper, the truncation scheme considered in 
Ref.~\cite{Watson:2011kv} is extended to the quark-antiquark \BS equation 
for pseudoscalar and vector mesons.  The primary aim of this study is to 
show how the infrared singular interaction may be consistently included 
into the formalism in such a way that physical quantities (meson masses 
and leptonic decay constants with arbitrary quark mass configurations) 
remain finite.  Given that both dynamical chiral symmetry breaking and 
the heavy quark limit are qualitatively (if not quantitatively) present 
in the quark propagator under this truncation, the corresponding signals 
for the meson spectrum will be investigated, both analytically and 
numerically.  As input, the interaction corresponding to a pure linear 
rising potential will be used, although it is recognized that this is an 
over-simplification from the physical standpoint.

The paper is organized as follows.  In the next section, we shall briefly 
review the quark propagator under the leading order Coulomb gauge 
truncation scheme.  Section~III then goes on to discuss the axialvector 
Ward-Takahashi identity within the context of Coulomb gauge and in the 
rest frame.  The \BS equation and leptonic decay constant for 
pseudoscalar mesons will then be discussed in Sec.~IV, with particular 
emphasis on how the infrared divergence of the interaction can be 
rendered finite.  In Sec.~V, the process will be repeated for vector 
mesons.  After briefly discussing the numerical implementation of the 
equations in Sec.~VI, the results will be presented in Sec.~VII.  
We finish with a summary and conclusions.

\section{Quark gap equation}
The central input into the \BS equation for quark-antiquark mesons stems 
from the quark propagator and the interaction between the quarks.  Let us 
thus begin by briefly reviewing some pertinent results for the quark 
propagator obtained previously under the leading order Coulomb gauge 
truncation scheme that will form the basis for this study 
\cite{Watson:2011kv}.  These results incorporate both dynamical chiral 
symmetry breaking and the heavy quark limit.

The leading order truncation scheme is based on the following 
instantaneous interaction term in the Coulomb gauge action:
\begin{equation}
{\cal S}\sim\int dx\,dy\,\left[
-\frac{1}{2}\rho_x^a\tilde{F}^{ab}(\vec{x},\vec{y})\de(x_0-y_0)\rho_y^b
\right],
\label{eq:sint}
\end{equation}
where $\rho^a$ is the color charge (superscript $a$ denotes here the 
color index in the adjoint representation) associated with the fields.  
In this case we are interested only in the quark component,
\begin{equation}
\rho_x^a=g\overline{q}_{x}\gamma^0T^aq_{x},
\end{equation}
with (conjugate) quark field at position $x$ denoted by 
($\overline{q}_x$) $q_x$, coupling $g$, and color matrix $T^a$ 
(Hermitian generator of the $SU(N_c)$ group).  In this study, we shall 
consider an interaction (color diagonal, and in momentum space) of the form
\begin{equation}
g^2C_F\tilde{F}(\vec{q}\,)
={\cal C}(2\pi)^3\de(\vec{q}\,)+\frac{8\pi\si}{(\vec{q\,}^2)^2}
\label{eq:int}
\end{equation}
($C_F=(N_c^2-1)/2N_c$ is the color factor associated with the quarks).  
Notice that since the interaction vanishes faster than $1/\vec{q\,}^2$ 
in the ultraviolet (UV), there will be no need to consider the 
renormalization of any quantities and we can omit any discussion of the 
renormalization constants.  The first term of the interaction arises from 
total color charge conservation 
(see Refs.~\cite{Reinhardt:2008pr,Watson:2011kv} for a detailed 
explanation): the constant ${\cal C}$ is considered finite until the end 
of the calculation, whereupon we take the limit 
${\cal C}\rightarrow\infty$.  Physically, ${\cal C}$ can be thought of 
as an infinite shift in the potential connected to the fact that one 
requires an infinite amount of energy to create a color nonsinglet, such 
as a single quark or gluon, from the vacuum (see later).  The second term 
is strongly infrared (IR) divergent with a coefficient $\si$.  The 
quark-antiquark potential, derived in the Coulomb gauge heavy quark 
limit and neglecting the contribution of pure Yang-Mills vertices 
\cite{Popovici:2010mb,Watson:2011kv} reads, for spatial separation $r$,
\begin{equation}
V(r)=g^2C_F\int\dk{\vec{q}}\tilde{F}(\vec{q}\,)
\left(1-e^{\imath\vec{q}\cdot\vec{r}}\right)=\si r,
\label{eq:pot}
\end{equation}
where $\dk{\vec{q}}=d\vec{q}/(2\pi)^3$.  The coefficient $\si$ can be 
identified as the string tension associated with a purely linear rising 
potential.  The specific value of $\si$ will not be of primary concern 
in this study because within the context of the leading order truncation 
scheme it will suffice that all dimensionfull quantities may be expressed 
in appropriate units of $\si$, such that the cancellation of the 
infrared singularities may be demonstrated.  As will be seen in Sec.~VII, 
the quantitative results arising from the above potential do not lend 
themselves to a detailed phenomenological study (this would require a 
more sophisticated truncation scheme beyond leading order, such as in 
Ref.~\cite{Pak:2011wu}).  However, we should specify some value for 
$\si$ in order to make a comparison to physical quantities.  The 
expectation value of $\tilde{F}$ is given by the instantaneous part of 
the temporal gluon propagator \cite{Cucchieri:2000hv,Watson:2011kv} and 
the string tension associated with the coefficient of the infrared 
divergence is called the Coulomb string tension ($\si_c$).  Thus, we 
should use $\si=\si_c$ as input.  It is known that $\si_c$ is equal to 
or larger than the physical Wilson string tension ($\si_W$) 
\cite{Zwanziger:2002sh} and lattice results indicate that $\si_c$ may be 
up to $\sim3$ times larger than $\si_W$ 
\cite{Iritani:2010mu,Voigt:2008rr,Nakagawa:2011ar} .  At the same time 
though, it will also be useful to compare with the earlier studies of 
Refs.~\cite{Govaerts:1983ft,Adler:1984ri,Alkofer:1988tc} where $\si_W$ 
was used.  In light of these considerations, we shall consider the range 
$\sqrt{\si}\sim440-762\mbox{MeV}$ (the lower value corresponding to 
$\sqrt{\si_W}$ and the upper value to $\sqrt{3\si_W}$) to give an 
estimate for the input scale.

Let us now turn to the quark propagator.  With the interaction, 
\eq{eq:sint}, and under the (leading order) truncation scheme to include 
only one loop terms, we have the following \DS equation 
(in momentum space) for the quark two-point proper function, 
$\G_{\ov{q}q}$ \cite{Watson:2011kv}:
\begin{align}
\G_{\ov{q}q}(p)&=\imath\left[\ga^0p_0-\s{\vec{\ga}}{\vec{p}}-m\right]
+\imath g^2C_F\int\dk{k}\tilde{F}(\vec{p}-\vec{k})
\ga^0W_{\ov{q}q}(k)\ga^0,
\label{eq:qdse}
\end{align}
where $m$ is the bare quark mass and $\dk{k}=d^4k/(2\pi)^4$.  
$\G_{\ov{q}q}$ can be written in terms of two scalar dressing functions 
$A_p$ and $B_p$ (the subscript $p$ denotes the spatial momentum 
dependence):
\begin{equation}
\G_{\ov{q}q}(p)
=\imath\left[\ga^0p_0-\s{\vec{\ga}}{\vec{p}}A_p-B_p\right].
\end{equation}
The corresponding quark propagator, $W_{\ov{q}q}(p)$ is
\begin{equation}
W_{\ov{q}q}(p)
=\frac{(-\imath)}{\Delta_f(p)}\left[
\ga^0p_0-\s{\vec{\ga}}{\vec{p}}A_p+B_p\right],\;\;\;\;
\Delta_f(p)=p_0^2-\vec{p\,}^2A_p^2-B_p^2+\imath 0_+.
\label{eq:qdec}
\end{equation}
Notice that the energy dependence of both the proper two-point function 
and the propagator is trivial.  This arises from the instantaneous 
character of the interaction and means that the dressing functions 
($A$ and $B$) are independent of the energy $p_0$.  This will lead to 
important simplifications in the \BS equation.  A possible fourth Dirac 
structure, proportional to $\ga^0p_0\vec{\ga}\cdot\vec{p}$ and arising 
in the noncovariant Coulomb gauge context does not appear, just as in the 
perturbative case \cite{Popovici:2008ty}.

The static quark propagator is defined as
\begin{equation}
W_{\ov{q}q}^{(s)}(\vec{p}\,)=\int\frac{dp^0}{2\pi}W_{\ov{q}q}(p)
=\frac{\s{\vec{\ga}}{\vec{p}}-M_p}{2\w_p},
\end{equation}
where we define the quark mass function, $M_p$, and quasiparticle energy, 
$\w_p$:
\be
M_p=\frac{B_p}{A_p},\;\;\;\;\w_p=\sqrt{\vec{p\,}^2+M_p^2}.
\ee
In the chiral limit, the quark condensate can be defined as 
(trace over Dirac matrices)
\be
\ev{\ov{q}q}=N_c\int\dk{\vec{p}}\mbox{Tr}_dW_{\ov{q}q}^{(s)}(\vec{p}\,)
=-2N_c\int\dk{\vec{p}}\frac{M_p}{\w_p}.
\label{eq:cond}
\ee
The meaning of the quark mass function becomes clear when one considers 
the quark \DS equation in the context of the interaction specified in 
\eq{eq:int}.  In terms of the dressing functions, $A$ and $B$, 
\eq{eq:qdse} decomposes to
\begin{align}
A_p&=1+\frac{1}{2}g^2C_F\int
\frac{\dk{\vec{k}}\tilde{F}(\vec{p}-\vec{k})}{\w_k}
\frac{\s{\vec{p}}{\vec{k}}}{\vec{p\,}^2},
\label{eq:aint}\\
B_p&=m+\frac{1}{2}g^2C_F\int
\frac{\dk{\vec{k}}\tilde{F}(\vec{p}-\vec{k})}{\w_k}M_k,
\label{eq:bint}
\end{align}
whereas combining the two equations in terms of the mass function leads 
to the Adler-Davis gap equation \cite{Adler:1984ri}:
\begin{equation}
M_p=m+\frac{1}{2}g^2C_F\int
\frac{\dk{\vec{k}}\tilde{F}(\vec{p}-\vec{k})}{\w_k}
\left[M_k-\frac{\s{\vec{p}}{\vec{k}}}{\vec{p\,}^2}M_p\right].
\label{eq:qgap}
\end{equation}
Given that the interaction $\tilde{F}$ contains both the divergent 
coefficient ${\cal C}$ multiplying the $\de$-function and the IR singular 
component, $A$ and $B$ are both divergent quantities.  However, the 
integrand of the gap equation contains exactly the cancellation necessary 
to ensure that the ratio of $A$ and $B$, namely, the quark mass function 
$M$, is finite.  The chiral condensate is also finite 
(see also Sec.~VII).  Considering the quark proper two-point function 
written in terms of the self-energy, $\Sigma$,
\begin{equation}
\G_{\ov{q}q}(p)=\imath\left[\ga^0p_0-\s{\vec{\ga}}{\vec{p}}-m_q\right]
+\Sigma(\vec{p}\,),
\end{equation}
the divergence inherent to the dressing functions $A$ and $B$ has the 
obvious interpretation of shifting the pole of the full quark propagator 
to infinity such that one would require infinite energy to excite a 
single quark from the vacuum.  The dynamical content of the quark 
propagator is contained within the finite mass function and we shall see 
that in the end, only the mass function (and associated quasiparticle 
energy) enters the \BS equation for color-singlet mesons.

\begin{figure}[t]
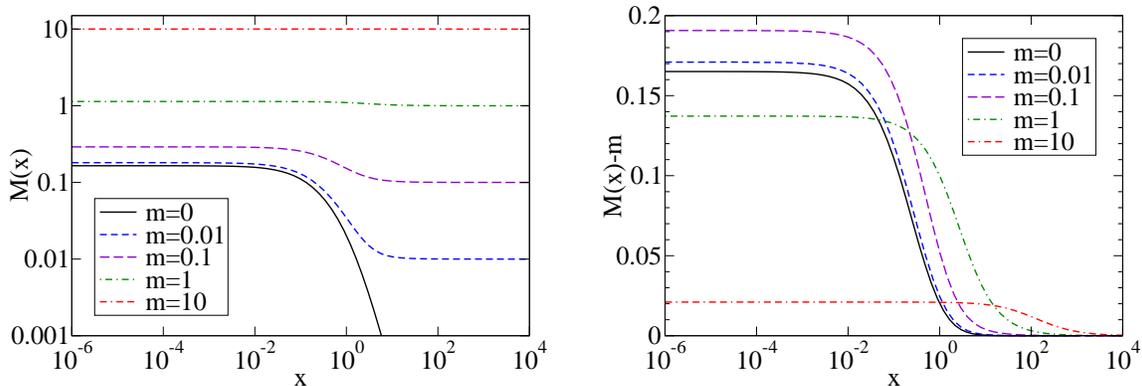

\vspace{0.8cm}
\includegraphics[width=0.4\linewidth]{eap0.eps}
\hspace{0.5cm}
\includegraphics[width=0.4\linewidth]{eap1.eps}
\vspace{0.3cm}
\caption{\label{fig:mfunc}[left panel] Quark mass function, $M(x)$, and 
[right panel] dressing, $M(x)-m$, plotted as functions of $x=\vec{p\,}^2$ 
for a range of quark masses.  All dimensionfull quantities are in 
appropriate units of the string tension, $\si$.  See text for details.}
\end{figure}

A detailed numerical solution to \eq{eq:qgap} for a range of quark masses 
was presented in Ref.~\cite{Watson:2011kv}.  Two aspects will be of 
relevance to this study and for the convenience of the reader, we briefly 
repeat them here.  The mass function is plotted (in appropriate units of 
the string tension, $\si$) in the left panel of Fig.~\ref{fig:mfunc}.  
One can see that in the chiral case ($m=0$) the quark mass function is 
nonzero, signaling the presence of dynamical chiral symmetry breaking.  
In the right panel of Fig.~\ref{fig:mfunc}, the dressing ($M-m$) is 
plotted.  As the bare quark mass, $m$, increases, the dressing initially 
increases but for larger $m$ starts to decrease.  In the (heavy quark) 
limit $m\rightarrow\infty$, one sees that $M\rightarrow m$.
The original derivation of \eq{eq:qgap} \cite{Adler:1984ri} arose from 
considering the chiral quark limit.  However, with the results for 
arbitrary quark masses it was shown \cite{Watson:2011kv} that the same 
equation reproduces the rest frame Coulomb gauge heavy quark limit (in 
the absence of pure Yang-Mills contributions) \cite{Popovici:2010mb}.  
Namely, the leading order truncation considered here encompasses both 
chiral and heavy quark physics.

\section{Axialvector Ward-Takahashi identity}
As is well understood, the pion can be regarded as the Goldstone boson of 
chiral symmetry breaking.  Considering a chiral rotation of the quark field,
\begin{equation}
q_x\rightarrow\exp{\left\{-\imath\al^a\frac{\ta^a}{2}\ga^5\right\}}q_x,
\end{equation}
where the $\al^a$ parametrize the rotation and the $\ta^a$ are the Pauli 
matrices corresponding to a two-flavor system ($a=1,2,3$), the pion vertex 
may be defined via the (chiral) quark self-energy as
\begin{equation}
\ta^a\G_{\pi}(\vec{p}\,)
=-\imath\frac{d}{d\al^a}\left[
\exp{\left\{\imath\al^b\frac{\ta^b}{2}\ga^5\right\}}
\Sigma(\vec{p}\,)
\exp{\left\{\imath\al^c\frac{\ta^c}{2}\ga^5\right\}}
\right]_{\al^a=0}
=\frac{\ta^a}{2}\left\{\ga^5,\Sigma(\vec{p}\,)\right\}.
\end{equation}
Such an approach was used to construct the pion \BS equation in, for 
example, Ref.~\cite{Govaerts:1983ft}.  More generally, the connection 
between the quark gap equation and the \BS equation for quark-antiquark 
bound states is expressed via the axialvector Ward-Takahashi identity 
(AXWTI).  In Coulomb gauge and in the chiral limit, the AXWTI was applied 
in, for example, Ref.~\cite{Adler:1984ri}.  Here, we are interested in 
finite, arbitrary mass quarks and the leptonic decay constants, so we 
shall closely follow the approach of Ref.~\cite{Maris:1997hd}.  There, 
the context was the consideration of covariant gauges.  Here, for the 
convenience of the reader, we shall repeat their argumentation for 
Coulomb gauge in the rest frame and specifically within our leading order 
truncation scheme.

Splitting the energy and spatial momentum arguments of the (noncovariant) 
propagators and proper functions, so that, for example, 
$\G_{\ov{q}q}(p)=\G_{\ov{q}q}(p^0,\vec{p}\,)$, let us consider the 
following combination (the energy and spatial momentum dependence of 
$\La$ will be explained below):
\begin{equation}
\La^{a5}(\vec{p}\,;p^0,P^0)
=\G_{\ov{q}q}^+(p^0+P^0/2,\vec{p}\,)\ga^5\frac{\ta^a}{2}
+\ga^5\frac{\ta^a}{2}\G_{\ov{q}q}^-(p^0-P^0/2,\vec{p}\,),
\label{eq:ldef}
\end{equation}
where the two quark proper two-point functions, $\G_{\ov{q}q}^\pm$, have 
arbitrary bare masses $m^\pm$ and correspondingly, different dressing 
functions $A^\pm$, $B^\pm$ and $M^\pm$.  Using the quark \DS equation, 
\eq{eq:qdse}, to rewrite the proper two-point functions, one obtains 
after some straightforward manipulation
\begin{align}
\La^{a5}(\vec{p}\,;p^0,P^0)
=&-\imath\ga^5\left[P^0\ga^0+(m^++m^-)\right]\frac{\ta^a}{2}\nonumber\\
&-\imath g^2C_F\int\dk{k}\tilde{F}(\vec{p}-\vec{k})
\ga^0W_{\ov{q}q}^+(k^0+P^0/2,\vec{k})\La^{a5}(\vec{k};k^0,P^0)
W_{\ov{q}q}^-(k^0-P^0/2,\vec{k})\ga^0.
\end{align}
We immediately notice that the above is a truncated \DS (or inhomogeneous 
Bethe-Salpeter) equation for a generalized quark-antiquark, color-singlet 
vertex $\La^{a5}$.  Given that the right-hand side expression is 
independent of the energy $p^0$, $\La^{a5}$ is also independent of $p^0$ 
and this arises because of the instantaneous character of the 
interaction.  The energy $P^0$ is the total energy flowing through the 
quark-antiquark pair, whereas the spatial momentum $\vec{p}$ 
(or $\vec{k}$ within the integral) flows along the quark line.  The 
vertex is thus in the rest frame of the quark-antiquark pair and this 
shall be extremely useful in our analysis.  It is convenient to use the 
shorthand notation $k^\pm$ to denote the energy and spatial momentum 
arguments $(k^0\pm P^0/2,\vec{k})$ (similarly for $p^\pm$) such that the 
above equation reads
\begin{equation}
\La^{a5}(\vec{p}\,;P^0)
=-\imath\ga^5\left[P^0\ga^0+(m^++m^-)\right]\frac{\ta^a}{2}
-\imath g^2C_F\int\dk{k}\tilde{F}(\vec{p}-\vec{k})
\ga^0W_{\ov{q}q}^+(k^+)\La^{a5}(\vec{k};P^0)W_{\ov{q}q}^-(k^-)\ga^0.
\label{eq:ldse}
\end{equation}
The axialvector Ward-Takahashi identity tells us that this generalized 
vertex can be rewritten as
\begin{equation}
\La^{a5}(\vec{p}\,;P^0)
=-\imath P^0\G_0^{a5}(\vec{p}\,;P^0)
-\imath(m^++m^-)\G^{a5}(\vec{p}\,;P^0),
\label{eq:axwti}
\end{equation}
where $\G_0^{a5}$ and $\G^{a5}$ are the temporal component of the 
axialvector vertex function and the pseudoscalar vertex function, 
respectively (both in the rest frame).  Separating the temporal 
axialvector and pseudoscalar components of $\La^{a5}$ in \eq{eq:ldse}, 
one can write down the \DS equations for $\G_0^{a5}$ and $\G^{a5}$:
\begin{align}
\G_0^{a5}(\vec{p}\,;P^0)=&
\ga^5\ga^0\frac{\ta^a}{2}
-\imath g^2C_F\int\dk{k}\tilde{F}(\vec{p}-\vec{k})
\ga^0W_{\ov{q}q}^+(k^+)\G_0^{a5}(\vec{k};P^0)W_{\ov{q}q}^-(k^-)\ga^0,
\label{eq:g50dse}\\
\G^{a5}(\vec{p}\,;P^0)=&
\ga^5\frac{\ta^a}{2}
-\imath g^2C_F\int\dk{k}\tilde{F}(\vec{p}-\vec{k})
\ga^0W_{\ov{q}q}^+(k^+)\G^{a5}(\vec{k};P^0)W_{\ov{q}q}^-(k^-)\ga^0.
\label{eq:g5dse}
\end{align}
Notice that both of these equations hold for arbitrary total energy 
$P^0$, i.e., they are not at resonance.

Now let us consider the homogeneous \BS equation in the pseudoscalar 
channel.  Taking \eq{eq:g5dse} and making the Ansatz that $\G^{a5}$ has 
a simple pole (in the rest frame and with the quantum numbers of a 
pseudoscalar vertex) with an as yet unknown residue, $r_{PS}$, of the form
\begin{equation}
\G^{a5}(\vec{p}\,;P^0)
=\frac{r_{PS}}{P_0^2-M_{PS}^2}\G_{PS}^{a}(\vec{p}\,;P^0)+\mbox{non-res.},
\label{eq:anszps}
\end{equation}
(with finite parts denoted `\mbox{non-res.}') then, at resonance, one 
has the homogeneous \BS equation:
\begin{equation}
\G_{PS}^{a}(\vec{p}\,;P^0)\stackrel{P_0^2=M_{PS}^2}{=}
-\imath g^2C_F\int\dk{k}\tilde{F}(\vec{p}-\vec{k})
\ga^0W_{\ov{q}q}^+(k^+)\G_{PS}^{a}(\vec{k};P^0)W_{\ov{q}q}^-(k^-)\ga^0.
\label{eq:psbse}
\end{equation}
Alternatively, taking \eq{eq:g50dse} and making instead the Ansatz that
\begin{equation}
\G_0^{a5}(\vec{p};P^0)
=\frac{r_{AV}P^0}{P_0^2-M_{PS}^2}\G_{PS}^{a}(\vec{p};P^0)+\mbox{non-res.},
\label{eq:anszav}
\end{equation}
one arrives at the identical homogeneous equation.  Combining 
Eqs.~(\ref{eq:anszps},\ref{eq:anszav}) and inserting into the AXWTI, 
\eq{eq:axwti}, one also has that
\begin{equation}
\La^{a5}(\vec{p}\,;P^0)
=-\imath\left[r_{AV}M_{PS}^2+r_{PS}(m^++m^-)\right]
\frac{\G_{PS}^{a}(\vec{p};P^0)}{P_0^2-M_{PS}^2}+\mbox{non-res.}
\end{equation}
Given the definition of $\La^{a5}$, \eq{eq:ldef}, and knowing that the 
quark proper two-point function contains no resonant components, the 
unknown residues $r_{PS}$ and $r_{AV}$ must obey the relation
\begin{equation}
r_{AV}M_{PS}^2=-r_{PS}(m^++m^-).
\label{eq:axwti1}
\end{equation}

\begin{figure}[t]
\begin{center}
\includegraphics[width=0.7\linewidth]{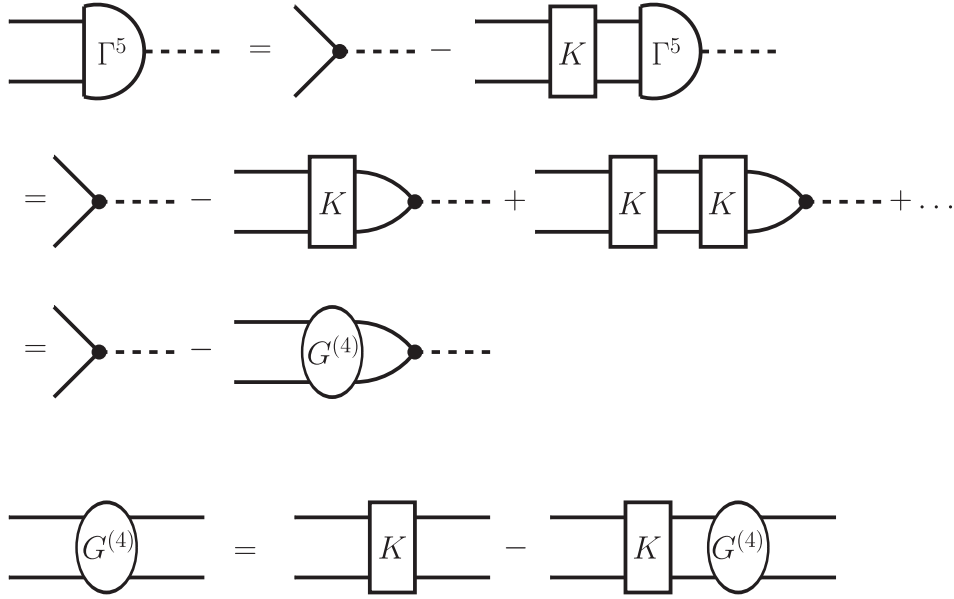}
\end{center}
\caption{\label{fig:kern}[top line] Diagrammatic representation of 
\eq{eq:g5dse1} in terms of the \BS kernel, $K$, [second line] expanded 
as a series in $K$ and [third line] the equivalent expression in terms 
of the amputated four-point function $G^{(4)}$.  [fourth line] The 
equation for $G^{(4)}$ (which can be also understood via the expansion 
in terms of $K$).  Internal lines denote dressed quark propagators, the 
dot denotes the tree-level vertex.  See text for details.}
\end{figure}

To obtain more information about the residues, some manipulation of the 
kernel is required.  Replacing the Dirac (and trivial flavor) indices and 
denoting the \BS kernel $K$ such that
\begin{equation}
K_{\al\ga;\de\ba}(\vec{p},\vec{k})
=\imath g^2C_F\tilde{F}(\vec{p}-\vec{k})
\left[\ga^0\right]_{\al\ga}\left[\ga^0\right]_{\de\ba},
\label{eq:bskern}
\end{equation}
\eq{eq:g5dse} can be written as
\begin{equation}
\G_{\al\ba}^{a5}(\vec{p}\,;P^0)=
\left[\ga^5\frac{\ta^a}{2}\right]_{\al\ba}
-\int\dk{k}K_{\al\ga;\de\ba}(\vec{p},\vec{k})
\left[
W_{\ov{q}q}^+(k^+)\G^{a5}(\vec{k};P^0)W_{\ov{q}q}^-(k^-)
\right]_{\ga\de},
\label{eq:g5dse1}
\end{equation}
with a similar form for \eq{eq:g50dse}.  Equation~(\ref{eq:g5dse1}) is 
represented diagrammatically in the top line of Fig.~\ref{fig:kern}.
The equation can be recursively expanded to an infinite series in the 
kernel as shown in the second line of Fig.~\ref{fig:kern}.  One sees 
that this has the effect of replacing the nonperturbative \BS vertex 
($\G^{a5}$) within the integral with the tree-level term ($\ga^5\ta^a/2$) 
and an infinite series in terms of the kernel ($K$).  With the form of 
the kernel under consideration here, \eq{eq:bskern}, this is nothing 
other than the ladder resummation of the \BS equation.  However, it is 
also known that the ladder resummation is directly connected to the 
amputated connected four-point quark-antiquark Green's function, $G^{(4)}$ 
(studied explicitly within the context of heavy quarks in the Coulomb 
gauge rest frame and with the same truncation scheme in 
Ref.~\cite{Popovici:2011yz}).  The equation for $G^{(4)}$ is shown 
diagrammatically in the last line of Fig.~\ref{fig:kern}.  Replacing then 
the infinite ladder series with $G^{(4)}$ in the equation for $\G^{a5}$ 
gives the third line of Fig.~\ref{fig:kern}.  
Equations (\ref{eq:g50dse},\ref{eq:g5dse}) can thus be rewritten as
\begin{align}
\G_{0\al\ba}^{a5}(\vec{p}\,;P^0)=&
\left[\ga^5\ga^0\frac{\ta^a}{2}\right]_{\al\ba}
-\int\dk{k} G_{\al\ba;\ga\de}^{(4)}(\vec{p},\vec{k};P^0)
\left[
W_{\ov{q}q}^+(k^+)\ga^5\ga^0\frac{\ta^a}{2}W_{\ov{q}q}^-(k^-)
\right]_{\de\ga},
\label{eq:g50dse2}\\
\G_{\al\ba}^{a5}(\vec{p}\,;P^0)=&
\left[\ga^5\frac{\ta^a}{2}\right]_{\al\ba}
-\int\dk{k} G_{\al\ba;\ga\de}^{(4)}(\vec{p},\vec{k};P^0)
\left[
W_{\ov{q}q}^+(k^+)\ga^5\frac{\ta^a}{2}W_{\ov{q}q}^-(k^-)
\right]_{\de\ga}.
\label{eq:g5dse2}
\end{align}
It is explicitly known that in the Coulomb gauge heavy quark limit, 
$G^{(4)}$ contains (physical) resonant components in the color-singlet 
channel \cite{Popovici:2011yz}.  In the pseudoscalar channel under 
consideration here, let us write
\be
G_{\al\ba;\ga\de}^{(4)}(\vec{p},\vec{k};P^0)
=\left[\G_{PS}^b(\vec{p};P^0)\right]_{\al\ba}
\frac{(-\imath N)}{P_0^2-M_{PS}^2}
\left[\ov{\G}_{PS}^b(\vec{k};-P^0)\right]_{\ga\de}+\mbox{non res.},
\label{eq:gres}
\ee
where $N$ is an unspecified normalization constant 
(we will use $N=1$ later).  The conjugate \BS vertex function is defined 
with the aid of the charge conjugation matrix $C$, such that
\be
\ov{\G}_{PS}^b(\vec{k};-P^0)=C\G_{PS}^{bT}(-\vec{k};-P^0)C^{-1}
\ee
($T$ denotes the transpose).  Taking the resonance Ansatz for 
$\G_{0}^{a5}$, \eq{eq:anszav}, or $\G^{a5}$, \eq{eq:anszps} as 
appropriate, along with \eq{eq:gres} for $G^{(4)}$, inserting into 
Eqs.~(\ref{eq:g50dse2},\ref{eq:g5dse2}), reorganizing the Dirac indices 
and evaluating the color and flavor traces, one obtains
\begin{align}
r_{AV}P^0&\stackrel{P_0^2=M_{PS}^2}{=}
-\imath NN_c\mbox{Tr}_d\int\dk{k}
\ga^5\ga^0W_{\ov{q}q}^+(k^+)\G_{PS}(\vec{k};P^0)W_{\ov{q}q}^-(k^-),
\nonumber\\
r_{PS}&\stackrel{P_0^2=M_{PS}^2}{=}
\imath NN_c\mbox{Tr}_d\int\dk{k}
\ga^5W_{\ov{q}q}^+(k^+)\G_{PS}(\vec{k};P^0)W_{\ov{q}q}^-(k^-).
\end{align}
The identity relating the residues, \eq{eq:axwti1}, can thus be written 
(independently of the normalization)
\begin{equation}
0\stackrel{P_0^2=M_{PS}^2}{=}{Tr}_d\int\dk{k}
\ga^5\left[P^0\ga^0-(m^++m^-)\right]
W_{\ov{q}q}^+(k^+)\G_{PS}(\vec{k};P^0)W_{\ov{q}q}^-(k^-).
\label{eq:axwti2}
\end{equation}

The leptonic decay constant for a pseudoscalar meson is defined as the 
coupling to the point axial field (see, for example, 
Refs.~\cite{Tandy:1997qf,Govaerts:1983ft,Maris:1997hd}).  In Minkowski 
space and in the rest frame, it can be defined as 
(trace over Dirac matrices)
\be
f_{PS}=\frac{N_c}{M_{PS}^2}\mbox{Tr}_{d}\int\dk{k}
\ga^5P^0\ga^0W_{\ov{q}q}^+(k^+)\G_{PS}^N(\vec{k};P^0)W_{\ov{q}q}^-(k^-)
\label{eq:fint}
\ee
where $\G_{PS}^N$ is the normalized \BS vertex function 
(the normalization will be discussed in the next section).  It is useful 
to define a second quantity
\begin{equation}
h_{PS}=N_c\mbox{Tr}_d\int\dk{k}
\ga^5W_{\ov{q}q}^+(k^+)\G_{PS}^N(\vec{k};P^0)W_{\ov{q}q}^-(k^-),
\label{eq:hint}
\end{equation}
such that \eq{eq:axwti2} reads
\begin{equation}
M_{PS}^2f_{PS}=(m^++m^-)h_{PS}.
\label{eq:fhcomp1}
\end{equation}
In the chiral limit, the Gell-Mann--Oakes--Renner (GMOR) relation 
\cite{Tandy:1997qf} states that
\begin{equation}
M_{PS}^2f_{PS}^2\stackrel{m^\pm\rightarrow0}{=}-(m^++m^-)\ev{\ov{q}q}.
\label{eq:gmor}
\end{equation}
Comparing with \eq{eq:fhcomp1}, we see that in the chiral limit
\begin{equation}
h_{PS}\stackrel{m^\pm\rightarrow0}{\longrightarrow}
-\frac{\ev{\ov{q}q}}{f_{PS}}=2N_c\int\dk{\vec{k}}\frac{M_k}{\w_kf_{PS}},
\label{eq:gmorh}
\end{equation}
showing that $h_{PS}$ is a generalization of the chiral condensate to 
finite quark masses.  Comparison to the GMOR relation will define one 
method of normalizing $\G_{PS}^N$ aside from the canonical normalization 
(to be discussed later).  Notice that with these conventions, one would 
have $f_{\pi}=92.2\mbox{MeV}$, a factor of $\sqrt{2}$ smaller than the 
convention used nowadays ($f_{\pi}=130.4\mbox{MeV}$ as defined in 
Ref.~\cite{Rosner:2012np}).

\section{Pseudoscalar mesons under truncation}
Having introduced the framework for studying the \BS equation in Coulomb 
gauge and under the leading order truncation, let us now show how this 
framework can be applied.  In this section, we shall be concerned with the 
pseudoscalar mesons; in the next section we shall repeat the analysis for 
vector mesons.  Our aim is to show how the homogeneous \BS equation for 
the meson mass, $M_{PS}$, the leptonic decay constant, $f_{PS}$, and the 
generalized condensate, $h_{PS}$, may be rewritten in terms of explicitly 
finite expressions despite the IR divergent interaction.

Removing the (overall) flavor factor, the general Dirac decomposition for 
the normalized pseudoscalar \BS vertex, at resonance and in the rest 
frame, can be written
\begin{equation}
\G_{PS}^N(\vec{p}\,;P^0)
=\ga^5\left[
\G_0^N(\vec{p\,}^2)+P^0\ga^0\G_1^N(\vec{p\,}^2)
+\s{\vec{\ga}}{\vec{p}}\,\G_2^N(\vec{p\,}^2)
+P^0\ga^0\s{\vec{\ga}}{\vec{p}}\,\G_3^N(\vec{p\,}^2)
\right].
\label{eq:pidec}
\end{equation}
In principle, the scalar dressing functions $\G_i^N$ have the arguments 
$\G_i^N(\vec{p}\,;P^0)$.  However in practice, since the homogeneous 
\BS equation used to derive these functions is valid only at resonance 
($P_0^2=M_{PS}^2$), the $\G_i^N$ are scalar functions of $\vec{p\,}^2$ 
alone and $P^0$ is merely a label.  To condense the notation, we shall 
write the momentum dependence as a subscript, i.e., $\G_{ip}^N$.  
Inserting the vertex decomposition, \eq{eq:pidec}, into the expressions 
for $f_{PS}$ and $h_{PS}$, Eqs.~(\ref{eq:fint}) and (\ref{eq:hint}), 
respectively, expanding the quark propagators using \eq{eq:qdec}, 
performing the Dirac traces and the energy integrals, one arrives at 
the following expressions:
\begin{align}
f_{PS}&=2\imath N_c\int
\frac{\dk{\vec{k}}}{\w_k^+\w_k^-
\left[P_0^2-(\tilde{\w}_k^++\tilde{\w}_k^-)^2\right]}\left\{
\G_{0k}^N\left[M_k^+\w_k^-+M_k^-\w_k^+\right]
+\G_{2k}^N\vec{k}^2\left[\w_k^+-\w_k^-\right]
\right.\nonumber\\&\left.
-(\tilde{\w}_k^++\tilde{\w}_k^-)
\frac{\left[M_k^++M_k^-\right]}{\left[\w_k^++\w_k^-\right]}
\left(\G_{1k}^N\left[M_k^+\w_k^-+M_k^-\w_k^+\right]
+\G_{3k}^N\vec{k}^2\left[\w_k^++\w_k^-\right]\right)
\right\},\nonumber\\
h_{PS}&=2\imath N_c\int
\frac{\dk{\vec{k}}}{\w_k^+\w_k^-
\left[P_0^2-(\tilde{\w}_k^++\tilde{\w}_k^-)^2\right]}\left\{
(\tilde{\w}_k^++\tilde{\w}_k^-)
\frac{\left[\w_k^++\w_k^-\right]}{\left[M_k^++M_k^-\right]}
\left(\G_{0k}^N\left[M_k^+\w_k^-+M_k^-\w_k^+\right]
+\G_{2k}^N\vec{k}^2\left[\w_k^+-\w_k^-\right]\right)
\right.\nonumber\\&\left.
-P_0^2\left(\G_{1k}^N\left[M_k^+\w_k^-+M_k^-\w_k^+\right]
+\G_{3k}^N\vec{k}^2\left[\w_k^++\w_k^-\right]\right)
\right\},
\end{align}
where we have introduced the notation
\begin{equation}
\tilde{\w}_k^\pm=A_k^\pm\w_k^\pm
\end{equation}
(in effect, the $\tilde{w}^\pm$ incorporate all the information about 
the IR divergence of the two quark self-energies and how they enter the 
expressions).  We notice that in the above, the following common 
combinations of functions arise:
\begin{align}
Q_{0k}^N&=\G_{0k}^N\left[M_k^+\w_k^-+M_k^-\w_k^+\right]
+\G_{2k}^N\vec{k}^2\left[\w_k^+-\w_k^-\right],\nonumber\\
Q_{1k}^N&=\G_{1k}^N\left[M_k^+\w_k^-+M_k^-\w_k^+\right]
+\G_{3k}^N\vec{k}^2\left[\w_k^++\w_k^-\right],
\label{eq:qdef}
\end{align}
allowing us to write
\begin{align}
f_{PS}&=2\imath N_c\int
\frac{\dk{\vec{k}}}{\w_k^+\w_k^-
\left[P_0^2-(\tilde{\w}_k^++\tilde{\w}_k^-)^2\right]}\left\{
Q_{0k}^N-(\tilde{\w}_k^++\tilde{\w}_k^-)
\frac{\left[M_k^++M_k^-\right]}{\left[\w_k^++\w_k^-\right]}Q_{1k}^N
\right\},\nonumber\\
h_{PS}&=2\imath N_c\int
\frac{\dk{\vec{k}}}{\w_k^+\w_k^-
\left[P_0^2-(\tilde{\w}_k^++\tilde{\w}_k^-)^2\right]}\left\{
(\tilde{\w}_k^++\tilde{\w}_k^-)
\frac{\left[\w_k^++\w_k^-\right]}{\left[M_k^++M_k^-\right]}
Q_{0k}^N-P_0^2Q_{1k}^N
\right\}.
\end{align}
As emphasized, one of the most important features of the above two 
quantities is that both are physical ($h_{PS}$ reduces, up to constant 
prefactors, to the chiral condensate in the chiral limit) and, in 
particular, cannot contain unphysical IR divergences such as those 
contained within the functions $\tilde{\w}^\pm$ (i.e., proportional 
to $A^\pm$).  As mentioned previously, in the Coulomb gauge heavy quark 
limit it is known that the amputated connected four-point 
quark-antiquark Green's function, $G^{(4)}$, explicitly contains 
physical resonance poles \cite{Popovici:2011yz}.  However the residues 
of these poles contain IR divergent factors, meaning that via the 
decomposition \eq{eq:gres}, we can expect the \BS vertex function 
$\G_{PS}^N$ and the components $\G_i^N$ to also involve IR divergent 
factors.  This will be made explicit shortly. In anticipation of the 
cancellations necessary so that the expressions for $f_{PS}$ and $h_{PS}$ 
are indeed finite, let us define the dimensionless functions
\begin{align}
f_k^N&
=\frac{1}{\left[P_0^2-(\tilde{\w}_k^++\tilde{\w}_k^-)^2\right]}\left\{
\frac{\left[\w_k^++\w_k^-\right]}{\left[M_k^++M_k^-\right]}Q_{0k}^N
-(\tilde{\w}_k^++\tilde{\w}_k^-)Q_{1k}^N
\right\},\nonumber\\
h_k^N&
=\frac{1}{\left[P_0^2-(\tilde{\w}_k^++\tilde{\w}_k^-)^2\right]}\left\{
\frac{(\tilde{\w}_k^++\tilde{\w}_k^-)}{\left[M_k^++M_k^-\right]}Q_{0k}^N
-\frac{P_0^2}{\left[\w_k^++\w_k^-\right]}Q_{1k}^N
\right\}.
\label{eq:fhdef}
\end{align}
This gives the expressions
\begin{align}
f_{PS}&=2\imath N_c\int\frac{\dk{\vec{k}}}{\w_k^+\w_k^-}
\frac{\left[M_k^++M_k^-\right]}{\left[\w_k^++\w_k^-\right]}f_k^N,
\label{eq:fpi}\\
h_{PS}&=2\imath N_c\int\frac{\dk{\vec{k}}}{\w_k^+\w_k^-}
\left[\w_k^++\w_k^-\right]h_k^N.
\label{eq:hpi}
\end{align}
Recall that in the chiral limit ($m^\pm\rightarrow0$), \eq{eq:gmorh} 
holds.  Comparing with \eq{eq:hpi}, we see that in terms of $h_k^N$ this 
is equivalent to
\begin{equation}
h_k^N\stackrel{m^\pm\rightarrow0}{\longrightarrow}
-\imath\frac{M_k}{2f_{PS}}.
\label{eq:hmpi}
\end{equation}
Given that $f_{PS}$ is a physical quantity and that the quark mass 
function, $M$, is IR finite, this limit tells us that indeed, $h_k^N$ 
should also be IR finite.  Further assuming a massless pion in the chiral 
limit and using the definitions Eqs.~(\ref{eq:qdef},\ref{eq:fhdef}), 
$h_k^N$ in the chiral limit can be written as
\begin{equation}
h_k^N\stackrel{m^\pm\rightarrow0}{\longrightarrow}
-\frac{1}{4\tilde{\w}_kM_k}Q_{0k}^N=-\frac{1}{2A_k}\G_{0k}^N,
\end{equation}
which would mean that the normalized \BS vertex function component 
$\G_0^N$ obeys
\begin{equation}
\G_{0k}^N\stackrel{m^\pm\rightarrow0}{\longrightarrow}
\imath\frac{B_k}{f_{PS}}.
\end{equation}
This result agrees explicitly with the standard result 
(see, for example, Ref.~\cite{Tandy:1997qf}).  Additionally, it confirms 
that the \BS vertex function and its components are IR divergent in 
exactly the same way as the quark self-energy.

Let us now turn to the homogeneous \BS equation in the pseudoscalar 
channel, \eq{eq:psbse}.  Inserting the decompositions, \eq{eq:qdec} for 
the quark propagators and \eq{eq:pidec} for the pseudoscalar \BS 
vertices, projecting out the Dirac components and performing the energy 
integrals one obtains four coupled homogeneous integral equations for 
the scalar functions $\G_i$ (un-normalized at this stage).  However, 
careful inspection reveals that remarkably, all four equations have 
integrands involving precisely the combinations of factors used in the 
definitions of the functions $f$ and $h$, 
Eqs.~(\ref{eq:qdef},\ref{eq:fhdef}).  The equations are:
\begin{align}
\G_{0p}&=-\frac{1}{2}g^2C_F\!\!
\int\frac{\dk{\vec{k}}\tilde{F}(\vec{p}-\vec{k})}{\w_k^+\w_k^-}
\left[\w_k^++\w_k^-\right]h_k
,\nonumber\\
\G_{1p}&=\frac{1}{2}g^2C_F\!\!
\int\frac{\dk{\vec{k}}\tilde{F}(\vec{p}-\vec{k})}{\w_k^+\w_k^-}
\frac{\left[M_k^++M_k^-\right]}{\left[\w_k^++\w_k^-\right]}f_k
,\nonumber\\
\G_{2p}&=-\frac{1}{2}g^2C_F\!\!
\int\frac{\dk{\vec{k}}\tilde{F}(\vec{p}-\vec{k})}{\w_k^+\w_k^-}
\frac{\s{\vec{p}}{\vec{k}}}{\vec{p\,}^2}
\frac{\left[M_k^+\w_k^--M_k^-\w_k^+\right]}{\vec{k}^2}h_k
,\nonumber\\
\G_{3p}&=\frac{1}{2}g^2C_F\!\!
\int\frac{\dk{\vec{k}}\tilde{F}(\vec{p}-\vec{k})}{\w_k^+\w_k^-}
\frac{\s{\vec{p}}{\vec{k}}}{\vec{p\,}^2}
\frac{\left[M_k^++M_k^-\right]}{\left[M_k^+\w_k^-+M_k^-\w_k^+\right]}f_k.
\label{eq:gbse}
\end{align}
The structure of the above set of equations shows that, in general, the 
\BS vertex with its four components is a convolution integral of the IR 
divergent interaction, the quark mass functions 
(and associated quasiparticle energies) and the two putatively IR finite 
functions $f$ and $h$.  Notice that for equal quarks, $\G_2$ vanishes.  
To arrive at a closed set of equations for $f$ and $h$, we must invert 
their definitions in terms of the $\G_i$.  Using 
Eqs.~(\ref{eq:qdef},\ref{eq:fhdef}), one may write
\begin{align}
\frac{P_0^2}{\left[\w_p^++\w_p^-\right]^2}f_p
-\frac{\left[\tilde{\w}_p^++\tilde{\w}_p^-\right]}
{\left[\w_p^++\w_p^-\right]}h_p
&=\G_{0p}\frac{\left[M_p^+\w_p^-+M_p^-\w_p^+\right]}
{\left[\w_p^++\w_p^-\right]\left[M_p^++M_p^-\right]}
+\G_{2p}\frac{\vec{p\,}^2\left[M_p^+-M_p^-\right]}
{\left[\w_p^++\w_p^-\right]^2},\nonumber\\
\frac{\left[\tilde{\w}_p^++\tilde{\w}_p^-\right]}
{\left[\w_p^++\w_p^-\right]}f_p-h_p
&=\G_{1p}\frac{\left[M_p^+\w_p^-+M_p^-\w_p^+\right]}
{\left[\w_p^++\w_p^-\right]}+\G_{3p}\vec{p\,}^2.
\label{eq:qinv}
\end{align}
Further, using the integral expression for the dressing function $A$, 
\eq{eq:aint}, one can rewrite the expression for $\tilde{\w}=A\w$ 
such that
\begin{equation}
\frac{\left[\tilde{\w}_p^++\tilde{\w}_p^-\right]}
{\left[\w_p^++\w_p^-\right]}
=1+\frac{1}{2}g^2C_F\int\frac{\dk{\vec{k}}\tilde{F}(\vec{p}-\vec{k})}
{\w_k^+\w_k^-}
\frac{\s{\vec{p}}{\vec{k}}}{\vec{p\,}^2}
\frac{\left[\w_p^+\w_k^-+\w_p^-\w_k^+\right]}
{\left[\w_p^++\w_p^-\right]}.
\label{eq:wdiv}
\end{equation}
Substituting this integral expression and the integral forms for the 
$\G_{i}$, \eq{eq:gbse}, into \eq{eq:qinv} one finds the final coupled 
equations for $f$ and $h$:
\begin{align}
h_p&=\frac{P_0^2}{\left[\w_p^++\w_p^-\right]^2}f_p
+\frac{1}{2}g^2C_F
\int\frac{\dk{\vec{k}}\tilde{F}(\vec{p}-\vec{k})}{\w_k^+\w_k^-}
\nonumber\\&\times
\left\{
h_k\frac{\left[\w_k^++\w_k^-\right]}{\left[\w_p^++\w_p^-\right]}\left(
\frac{\left[M_p^+\w_p^-+M_p^-\w_p^+\right]}{\left[M_p^++M_p^-\right]}
+\frac{\s{\vec{p}}{\vec{k}}}{\vec{k}^2}
\frac{\left[M_k^+\w_k^--M_k^-\w_k^+\right]\left[M_p^+-M_p^-\right]}
{\left[\w_k^++\w_k^-\right]\left[\w_p^++\w_p^-\right]}
\right)
-h_p\frac{\s{\vec{p}}{\vec{k}}}{\vec{p\,}^2}
\frac{\left[\w_p^+\w_k^-+\w_p^-\w_k^+\right]}
{\left[\w_p^++\w_p^-\right]}
\right\},\nonumber\\
f_p&=h_p+\frac{1}{2}g^2C_F
\int\frac{\dk{\vec{k}}\tilde{F}(\vec{p}-\vec{k})}{\w_k^+\w_k^-}
\nonumber\\&\times
\left\{
f_k\frac{\left[M_k^++M_k^-\right]}{\left[\w_k^++\w_k^-\right]}\left(
\frac{\left[M_p^+\w_p^-+M_p^-\w_p^+\right]}{\left[\w_p^++\w_p^-\right]}
+\frac{\s{\vec{p}}{\vec{k}}\left[\w_k^++\w_k^-\right]}
{\left[M_k^+\w_k^-+M_k^-\w_k^+\right]}
\right)
-f_p\frac{\s{\vec{p}}{\vec{k}}}{\vec{p\,}^2}
\frac{\left[\w_p^+\w_k^-+\w_p^-\w_k^+\right]}
{\left[\w_p^++\w_p^-\right]}
\right\}.
\label{eq:psbses}
\end{align}
In both equations, the divergence (arising either from the $\de$-function 
charge constraint proportional to the divergent constant ${\cal C}$ or 
from the infrared singular $1/\vec{q\,}^4$ term in $\tilde{F}$) occurs 
when $\vec{p}=\vec{k}$.  It is straightforward to show that in both 
cases, the remaining factors of the integrand cancel.  Thus, the above 
integral equations are explicitly infrared finite and so are the 
functions $f$ and $h$, as demanded by the definitions of $f_{PS}$ and 
$h_{PS}$.  Notice that if the meson were anything but a color-singlet, 
the cancellation of the divergences would not occur, just as in the case 
for the heavy quark limit 
\cite{Popovici:2010mb,Popovici:2010ph,Popovici:2011yz}.  Further, we see 
that the only occurrence of $P_0^2=M_{PS}^2$ lies in the coupling of $f$ 
to the $h$-equation.  That the system of four IR divergent components of 
the general pseudoscalar \BS vertex decomposition reduces to a system of 
two finite functions is analogous to the case of the quark propagator, 
where the two IR divergent dressing functions ($A$ and $B$) conspire to 
form the finite gap equation for the mass function, $M$.

In the special case where the quarks have equal mass, the above equations 
reduce significantly, although their general form remains the same.  The 
equal mass equations read
\begin{align}
h_p&=\frac{P_0^2}{4\w_p^2}f_p
+\frac{1}{2}g^2C_F\int\frac{\dk{\vec{k}}\tilde{F}(\vec{p}-\vec{k})}{\w_k}
\left\{h_k-h_p\frac{\s{\vec{p}}{\vec{k}}}{\vec{p\,}^2}\right\},
\nonumber\\
f_p&=h_p
+\frac{1}{2}g^2C_F\int\frac{\dk{\vec{k}}\tilde{F}(\vec{p}-\vec{k})}{\w_k}
\left\{
f_k\frac{\left[\s{\vec{p}}{\vec{k}}+M_pM_k\right]}
{\left[\vec{k}^2+M_k^2\right]}
-f_p\frac{\s{\vec{p}}{\vec{k}}}{\vec{p\,}^2}
\right\}.
\label{eq:psbsee}
\end{align}
Comparing the first equation with the gap equation for the quark mass 
function, \eq{eq:qgap}, one sees that the integral has the same form but 
with $M$ replaced with $h$.  Thus, it is clear that in the chiral limit, 
$h$ is proportional to $M$ (as shown in the discussion leading up to 
\eq{eq:hmpi}) and $P_0^2=M_{PS}^2=0$, showing that the pion is massless 
in the chiral limit, as it should be.

It is worth pointing out a major difference between the \BS equation 
above and studies performed within the covariant approach, e.g., 
Refs.~\cite{Maris:1997tm,Maris:1999nt,Alkofer:2002bp}.  In the covariant 
gauge approach, one works in Euclidean space and since the bound state 
total momentum is timelike, it is necessary to make an extension of the 
quark propagator to complex Euclidean momenta (usually numerically).  
This extension into the complex plane has the consequence that possible 
singularities must be treated carefully or avoided altogether 
(although numerical techniques to deal with such issues have been 
developed, e.g., 
Refs.~\cite{Bhagwat:2002tx,Fischer:2005en,
Krassnigg:2009gd,Fischer:2008sp}).  For mesons of roughly equal quark 
masses this is not a major issue, however, for asymmetric systems such 
as heavy-light mesons this becomes challenging.  In the Coulomb gauge 
rest frame (and Minkowski space) equations presented above, only the 
quark mass function (and quasiparticle energy) enters and is always 
evaluated at spatial momenta, rendering the issue of continuation to 
timelike momenta irrelevant (effectively, the energy integrals 
performed analytically in the derivation of the equations deal with 
this issue from the outset).  It will be seen that heavy-light systems 
are no more difficult to treat than equal quark systems.

In order to calculate $f_{PS}$ and $h_{PS}$, we must normalize the 
functions $f$ and $h$.  As a first derivation, let us consider the equal 
mass case where
\begin{equation}
f_{PS}=2\imath N_c\int\frac{\dk{\vec{k}}M_k}{\w_k^3}f_k^N.
\end{equation}
However, in the chiral limit, \eq{eq:hmpi} holds such that we can write
\begin{equation}
M_k\stackrel{m\rightarrow0}{\longrightarrow}2\imath f_{PS} h_k^N,
\end{equation}
which would mean that the normalized functions must obey
\begin{equation}
1\stackrel{m\rightarrow0}{=}
-4N_c\int\frac{\dk{\vec{k}}}{\w_k^3}f_k^Nh_k^N
\label{eq:norm0}
\end{equation}
in this limit, such that the GMOR relation, \eq{eq:gmor}, is satisfied.  
We see that in this restricted limit, the normalization can be written 
in terms of our finite functions $f$ and $h$.  More properly, the 
canonical normalization of the \BS vertex function reads 
(in the Minkowski space rest frame and with trace over color, flavor 
and Dirac matrices) \cite{Tandy:1997qf}
\begin{equation}
2\imath\de^{ab}P^0=\mbox{Tr}_{f,c,d}\int\!\!\dk{k}\!\!
\left\{\ov{\G}_{PS}^{aN}(\vec{k};-P^0)
\frac{\pd W_{\ov{q}q}^+(k^+)}{\pd P^0}\G_{PS}^{bN}(\vec{k};P^0)
W_{\ov{q}q}^-(k^-)+\ov{\G}_{PS}^{aN}(\vec{k};-P^0)W_{\ov{q}q}^+(k^+)
\G_{PS}^{bN}(\vec{k};P^0)\frac{\pd W_{\ov{q}q}^-(k^-)}{\pd P^0}\right\}.
\end{equation}
Evaluating the flavor (recall that we have $N_f=2$ and consider flavor 
nonsinglet mesons) and color traces, and using
\begin{equation}
\frac{\pd W_{\ov{q}q}^\pm(k^\pm)}{\pd P^0}
=-W_{\ov{q}q}^\pm(k^\pm)
\frac{\pd\G_{\ov{q}q}^\pm(k^\pm)}{\pd P^0}
W_{\ov{q}q}^\pm(k^\pm),\;\;\;\;
\frac{\pd\G_{\ov{q}q}^\pm(k^\pm)}{\pd P^0}=\pm\frac{\imath}{2}\ga^0,
\end{equation}
one obtains
\begin{equation}
1=-\frac{N_c}{2P^0}\mbox{Tr}_d\int\dk{k}
\ov{\G}_{PS}^N(\vec{k};-P^0)W_{\ov{q}q}^+(k^+)
\left\{
\ga^0W_{\ov{q}q}^+(k^+)\G_{PS}^N(\vec{k};P^0)
-\G_{PS}^N(\vec{k};P^0)W_{\ov{q}q}^-(k^-)\ga^0
\right\}W_{\ov{q}q}^-(k^-).
\end{equation}
The (Dirac structure of the) conjugate \BS vertex function is explicitly 
given in the rest frame by
\begin{equation}
\ov{\G}^N(\vec{k};-P^0)=C\G^{NT}(-\vec{k};-P^0)C^{-1}
=\ga^5\left[
\G_{0k}^N-P^0\ga^0\G_{1k}^N-\s{\vec{\ga}}{\vec{k}}\G_{2k}^N
-P^0\ga^0\s{\vec{\ga}}{\vec{k}}\G_{3k}^N
\right].
\end{equation}
After evaluating the Dirac trace and performing the energy integrals, 
one finds after some effort that the normalization condition can be 
written in terms of the finite functions $f$ and $h$:
\begin{equation}
1=-4N_c\int\frac{\dk{\vec{k}}f_k^Nh_k^N}{\w_k^+\w_k^-}
\frac{\left[M_k^++M_k^-\right]}{\left[M_k^+\w_k^-+M_k^-\w_k^+\right]},
\label{eq:norm}
\end{equation}
which reduces explicitly for equal mass quarks to the earlier form 
\eq{eq:norm0}.  Given an arbitrarily normalized solution to the \BS 
equation, \eq{eq:psbses}, (we shall use the condition 
$h(\vec{p\,}^2\rightarrow0)=1$ for the numerical analysis later on) it 
is useful to write
\begin{equation}
f_k^N
=\frac{f_k}{\imath{\cal N}},\;\;\;\;h_k^N=\frac{h_k}{\imath{\cal N}}.
\label{eq:normn}
\end{equation}
This then results in the normalization condition
\begin{equation}
{\cal N}^2=4N_c\int\frac{\dk{\vec{k}}f_kh_k}{\w_k^+\w_k^-}
\frac{\left[M_k^++M_k^-\right]}{\left[M_k^+\w_k^-+M_k^-\w_k^+\right]}
\end{equation}
and the redefinition of $f_{PS}$ and $h_{PS}$ to
\begin{align}
f_{PS}&=\frac{2N_c}{{\cal N}}\int\frac{\dk{\vec{k}}}{\w_k^+\w_k^-}
\frac{\left[M_k^++M_k^-\right]}{\left[\w_k^++\w_k^-\right]}f_k,
\\
h_{PS}&=\frac{2N_c}{{\cal N}}\int\frac{\dk{\vec{k}}}{\w_k^+\w_k^-}
\left[\w_k^++\w_k^-\right]h_k.
\end{align}

Recall that it is known that the Coulomb gauge quark propagator under 
this truncation reduces to its heavy quark limit (in the absence of pure 
Yang-Mills theory corrections) \cite{Watson:2011kv}.  The difference 
between the mass function and the bare mass becomes smaller in the heavy 
quark limit, i.e., $M_k-m_h\rightarrow0$ as $m_h\rightarrow\infty$ 
(we will use $m_h$ to denote the heavy quark mass).  For a heavy-light 
system, with $m^+=m_h\rightarrow\infty$ and assuming that the above 
integrals for ${\cal N}$, $f_{PS}$ and $h_{PS}$ converge for large 
$\vec{k}$ (i.e., that the convergence is driven by the large $\vec{k}$ 
behavior of the functions $f_k$ and $h_k$) then we can make the 
replacement $\w_k^+=M_k^+=m_h\gg\w_k^-,M_k^-$ within the integrals.  
This then gives
\begin{equation}
{\cal N}^2\stackrel{m^+=m_h}{\longrightarrow}
\frac{4N_c}{m_h}\int
\frac{\dk{\vec{k}}f_kh_k}{\w_k^-\left[\w_k^-+M_k^-\right]},\;\;
f_{PS}\stackrel{m^+=m_h}{\longrightarrow}
\frac{2N_c}{{\cal N}m_h}\int\frac{\dk{\vec{k}}}{\w_k^-}f_k,\;\;
h_{PS}\stackrel{m^+=m_h}{\longrightarrow}
\frac{2N_c}{{\cal N}}\int\frac{\dk{\vec{k}}}{\w_k^-}h_k.
\end{equation}
Performing the same substitution for \eq{eq:psbses} gives
\begin{equation}
h_p\stackrel{m^+=m_h}{\longrightarrow}
\frac{P_0^2}{m_h^2}f_p+{\cal O}(1/m_h),\;\;\;\;
f_p\stackrel{m^+=m_h}{\longrightarrow}h_p+{\cal O}(1/m_h),
\end{equation}
which tells us that (asymptotically, as $m_h\rightarrow\infty$) 
$P_0^2=M_{PS}^2=m_h^2$ as we would expect.  We notice though, that the 
leading order heavy quark limit merely tells us that $f_p=h_p$, but not 
what $f_p$ or $h_p$ actually are.  With the above limits for $M_{PS}$, 
$f_{PS}$ and $h_{PS}$, we see that the relationship \eq{eq:fhcomp1} is 
explicitly fulfilled.  The most important result of this analysis though, 
is that as $m_h\rightarrow\infty$ 
(assuming the integrals are well-defined) and given 
${\cal N}\sim1/\sqrt{m_h}$,
\begin{equation}
f_{PS}\sqrt{M_{PS}}\sim\mbox{const.}
\label{eq:hlps}
\end{equation}
This result explicitly agrees with the analysis of heavy quark effective 
theory at leading order \cite{Neubert:1993mb}.

At this stage, it is appropriate to make a comparison of the formalism 
presented here with similar works on the Coulomb gauge \BS equation.  As 
mentioned earlier, the chiral limit pseudoscalar meson (i.e., the pion) 
was investigated in Ref.~\cite{Govaerts:1983ft}.  An extension to the 
case of equal, but finite mass quarks was performed in, e.g., 
Refs.~\cite{Alkofer:1988tc,Langfeld:1989en,Alkofer:2005ug}.  The 
equations therein have subtle differences to those presented here, over 
and above the obvious extension to arbitrary quark masses.  Since the 
referenced works dealt with equal mass quarks, the vertex component 
$\G_2$ was absent from the outset although they explicitly found that 
the system can be decomposed in terms of two finite functions.  The 
coupled equations for the two finite functions ($f$ and $h$) presented 
here are, from the point of view of canceling the IR divergences, 
somewhat more explicit than the earlier versions since the IR divergent 
factors ($\tilde{\w}$ in our notation) inherent to the quark self-energy 
have been completely eliminated.  This elimination also extends to the 
equations for $f_{PS}$, $h_{PS}$ and the normalization.  The success of 
making this cancellation explicit arises from utilizing a different 
definition of the two functions, \eq{eq:fhdef}, designed with the 
arbitrary quark mass expressions for $f_{PS}$ and $h_{PS}$ in mind.  This 
is in distinction to the original decomposition of 
Refs.~\cite{Govaerts:1983ft,Alkofer:1988tc}, whose motivation was to 
make the \BS equation tractable.  Despite the differences in the 
decomposition, both sets of functions yield finite quantities.

\section{Vector mesons under truncation}
Let us now consider vector mesons.  Although the expressions are 
considerably more involved than those discussed in the previous section, 
they follow the same pattern.  The homogeneous \BS equation retains its 
form as for the pseudoscalar case, \eq{eq:psbse},
\begin{equation}
\G_{Vi}^{a}(\vec{p};P^0)\stackrel{P_0^2=M_{V}^2}{=}
-\imath g^2C_F\int\dk{k}\tilde{F}(\vec{p}-\vec{k})
\ga^0W_{\ov{q}q}^+(k^+)\G_{Vi}^{a}(\vec{k};P^0)W_{\ov{q}q}^-(k^-)\ga^0,
\label{eq:vbse}
\end{equation}
but the general decomposition of the normalized vector \BS vertex 
function (in the rest frame and omitting the flavor dependence) now reads
\begin{align}
\G_{Vi}^{N}(\vec{p};P^0)=&\left\{
\ga^i\left[\G_{0p}^{N}+P^0\ga^0\G_{1p}^{N}
+\s{\vec{\ga}}{\vec{p}}\,\G_{2p}^{N}
+P^0\ga^0\s{\vec{\ga}}{\vec{p}}\,\G_3^{N}\right]
\right.\nonumber\\&\left.
+p^i\left[\G_{2p}^{N}+\G_{4p}^{N}+P^0\ga^0(\G_{5p}^{N}
-\G_{3p}^{N})+\s{\vec{\ga}}{\vec{p}}\,\G_{6p}^{N}
+P^0\ga^0\s{\vec{\ga}}{\vec{p}}\,\G_{7p}^{N}\right]\right\},
\label{eq:vdec}
\end{align}
where it is recognized that the vector meson at resonance is transverse 
to the total four-momentum $P$ and so, can only have spatial components.  
The vector meson leptonic decay constant can be defined as 
(trace over Dirac matrices) \cite{Maris:1999nt}
\begin{equation}
f_V=-\frac{N_c}{3M_V}\mbox{Tr}_{d}\int\dk{k}
\ga^iW_{\ov{q}q}^+(k^+)\G_{Vi}^N(\vec{k};P^0)W_{\ov{q}q}^-(k^-)
\label{eq:fv0}
\end{equation}
and the normalization defined as 
(trace over color, flavor and Dirac matrices)
\begin{equation}
6\imath\de^{ab}P^0=\mbox{Tr}_{f,c,d}\int\!\!\dk{k}\!\!
\left\{\ov{\G}_{Vi}^{aN}(\vec{k};-P^0)
\frac{\pd W_{\ov{q}q}^+(k^+)}{\pd P^0}\G_{Vi}^{bN}(\vec{k};P^0)
W_{\ov{q}q}^-(k^-)
+\ov{\G}_{Vi}^{aN}(\vec{k};-P^0)W_{\ov{q}q}^+(k^+)
\G_{Vi}^{bN}(\vec{k};P^0)\frac{\pd W_{\ov{q}q}^-(k^-)}{\pd P^0}\right\}.
\label{eq:vnorm}
\end{equation}
In both the latter equations, a factor of three arises from the 
(transverse) polarization states of the vector meson.  With the 
definition of the conjugate amplitude as before:
\begin{equation}
\ov{\G}_{Vi}^N(\vec{k};-P^0)=C\G_{Vi}^{NT}(-\vec{k};-P^0)C^{-1},
\end{equation}
one has
\begin{align}
\ov{\G}_{Vi}^N(\vec{k};-P^0)=&
\ga^i\left[-\G_{0k}^N+P^0\ga^0\G_{1k}^N
+\s{\vec{\ga}}{\vec{k}}\G_{2k}^N
+P^0\ga^0\s{\vec{\ga}}{\vec{k}}\G_{3k}^N\right]
\nonumber\\&
+k^i\left[\G_{2k}^N-\G_{4k}^N-P^0\ga^0(\G_{5k}^N+\G_{3k}^N)
-\s{\vec{\ga}}{\vec{k}}\G_{6k}^N
+P^0\ga^0\s{\vec{\ga}}{\vec{k}}\G_{7k}^N\right].
\label{eq:cvdec}
\end{align}

Starting with the leptonic decay constant, $f_V$, defined in 
\eq{eq:fv0}, after inserting the decompositions for the \BS vertex, 
\eq{eq:vdec}, and the quark propagator, \eq{eq:qdec}, evaluating the 
trace over Dirac matrices and performing the energy integrals, one has 
the expression
\begin{align}
f_V=&-\frac{2\imath N_c}{3M_V}\int
\frac{\dk{\vec{k}}}
{\w_k^+\w_k^-\left[P_0^2-(\tilde{\w}_k^++\tilde{\w}_k^-)^2\right]}
\nonumber\\&\times\left\{
(\tilde{\w}_k^++\tilde{\w}_k^-)
\left[-2\frac{\left[\w_k^++\w_k^-\right]}{\left[M_k^++M_k^-\right]}
\left(\G_{0k}^N\left[M_k^+\w_k^-+M_k^-\w_k^+\right]
+\G_{2k}^N\vec{k}^2\left[\w_k^+-\w_k^-\right]\right)
\right.\right.\nonumber\\&\left.\left.
-\frac{\left[M_k^++M_k^-\right]}{\left[\w_k^++\w_k^-\right]}
\left(\left[\G_{0k}^N+\vec{k}^2\G_{6k}^N\right]
\left[M_k^+\w_k^-+M_k^-\w_k^+\right]
-\G_{4k}^N\vec{k}^2\left[\w_k^++\w_k^-\right]\right)\right]
\right.\nonumber\\&\left.
+P_0^2\left[2\G_{1k}^N\left[M_k^+\w_k^-+M_k^-\w_k^+\right]
+2\G_{3k}^N\vec{k}^2\left[\w_k^++\w_k^-\right]
+\left[\G_{1k}^N-\vec{k}^2\G_{7k}^N\right]
\left[M_k^+\w_k^-+M_k^-\w_k^+\right]
+\G_{5k}^N\vec{k}^2\left[\w_k^+-\w_k^-\right]\right]
\right\}.\nonumber\\
\end{align}
Interestingly, we note the appearance of the combinations of functions 
$Q_0^N$ and $Q_1^N$, \eq{eq:qdef}, defined for the pseudoscalar meson.  
Introducing the further two combinations
\begin{align}
Q_{2k}^N=&
\left[\G_{0k}^N+\vec{k}^2\G_{6k}^N\right]
\left[M_k^+\w_k^-+M_k^-\w_k^+\right]
-\G_{4k}^N\vec{k}^2\left[\w_k^++\w_k^-\right],\nonumber\\
Q_{3k}^N=&
\left[\G_{1k}^N-\vec{k}^2\G_{7k}^N\right]
\left[M_k^+\w_k^-+M_k^-\w_k^+\right]
+\G_{5k}^N\vec{k}^2\left[\w_k^+-\w_k^-\right],
\end{align}
we can write
\begin{align}
f_V=\frac{2\imath N_c}{3M_V}&
\int\frac{\dk{\vec{k}}}{\w_k^+\w_k^-
\left[P_0^2-(\tilde{\w}_k^++\tilde{\w}_k^-)^2\right]}
\nonumber\\&\times
\left\{
(\tilde{\w}_k^++\tilde{\w}_k^-)
\left[
2\frac{\left[\w_k^++\w_k^-\right]}{\left[M_k^++M_k^-\right]}Q_{0k}^N
+\frac{\left[M_k^++M_k^-\right]}{\left[\w_k^++\w_k^-\right]}Q_{2k}^N
\right]
-P_0^2\left[2Q_{1k}^N+Q_{3k}^N\right]
\right\}.
\end{align}
We identify the appearance of $h_k^N$ (defined in \eq{eq:fhdef} for the 
pseudoscalar meson) in this equation and introduce, in analogy to before, 
the dimensionless functions
\begin{align}
g_k^N=&
\frac{1}{\left[P_0^2-(\tilde{\w}_k^++\tilde{\w}_k^-)^2\right]}
\left\{Q_{2k}^N
-\frac{(\tilde{\w}_k^++\tilde{\w}_k^-)\left[\w_k^++\w_k^-\right]}
{\left[M_k^++M_k^-\right]}Q_{3k}^N\right\},\nonumber\\
j_k^N=&
\frac{1}{\left[P_0^2-(\tilde{\w}_k^++\tilde{\w}_k^-)^2\right]}
\left\{
\frac{(\tilde{\w}_k^++\tilde{\w}_k^-)}
{\left[\w_k^++\w_k^-\right]}Q_{2k}^N
-\frac{P_0^2}{\left[M_k^++M_k^-\right]}Q_{3k}^N\right\},
\label{eq:gjdef}
\end{align}
where the function $g_k^N$ will shortly be needed and the aim will be to 
see if these functions are finite.  The vector meson leptonic decay 
constant can thus be written as
\begin{equation}
f_V=\frac{2\imath N_c}{3M_V}\int\frac{\dk{\vec{k}}}{\w_k^+\w_k^-}
\left(
2\left[\w_k^++\w_k^-\right]h_k^N+\left[M_k^++M_k^-\right]j_k^N
\right).
\label{eq:fv}
\end{equation}
The above expression has an almost identical form as for the pseudoscalar 
case, \eq{eq:fpi}.

Turning now to the vector meson \BS equation, \eq{eq:vbse}, after 
inserting the decompositions for the \BS vertex, \eq{eq:vdec}, and quark 
propagator, \eq{eq:qdec}, projecting out the Dirac structure and 
performing the energy integrals, one can identify the appearance of 
precisely the combinations of (un-normalized) functions that define 
$f_k$, $h_k$, $g_k$ and $j_k$, Eqs.~(\ref{eq:fhdef},\ref{eq:gjdef}), 
(the former two defined for the pseudoscalar meson) in exactly the same 
way as previously.  That there are only four combinations means that 
there are only four independent dynamical functions that characterize 
the vector meson under this truncation.  The equations for the vector 
meson $\G_i$ read:
\begin{align}
\G_{0p}=&\frac{1}{2}g^2C_F\int
\frac{\dk{\vec{k}}\tilde{F}(\vec{p}-\vec{k})}{\w_k^+\w_k^-}
\left\{
\left(\frac{\Delta}{2}-1\right)
\left[\w_k^++\w_k^-\right]h_k
-\frac{\Delta}{2}\left[M_k^++M_k^-\right]j_k\right\},\nonumber\\
\G_{1p}=&\frac{1}{2}g^2C_F\int
\frac{\dk{\vec{k}}\tilde{F}(\vec{p}-\vec{k})}{\w_k^+\w_k^-}
\left\{\left(1-\frac{\Delta}{2}\right)
\frac{\left[M_k^++M_k^-\right]}{\left[\w_k^++\w_k^-\right]}f_k
+\frac{\Delta}{2}g_k\right\},\nonumber\\
\G_{2p}=&\frac{1}{2}g^2C_F\int
\frac{\dk{\vec{k}}\tilde{F}(\vec{p}-\vec{k})}{\w_k^+\w_k^-}
\frac{\s{\vec{p}}{\vec{k}}}{\vec{p\,}^2\vec{k}^2}
\left[M_k^+\w_k^--M_k^-\w_k^+\right]\left[-h_k\right],\nonumber\\
\G_{3p}=&\frac{1}{2}g^2C_F\int
\frac{\dk{\vec{k}}\tilde{F}(\vec{p}-\vec{k})}{\w_k^+\w_k^-}
\frac{\s{\vec{p}}{\vec{k}}}{\vec{p\,}^2}
\frac{\left[M_k^++M_k^-\right]}{\left[M_k^+\w_k^-+M_k^-\w_k^+\right]}f_k,
\nonumber\\
\G_{4p}=&\frac{1}{2}g^2C_F\int
\frac{\dk{\vec{k}}\tilde{F}(\vec{p}-\vec{k})}{\w_k^+\w_k^-}
\frac{\s{\vec{p}}{\vec{k}}}{\vec{p\,}^2}
\frac{\left[\w_k^++\w_k^-\right]\left[M_k^++M_k^-\right]}
{\left[M_k^+\w_k^-+M_k^-\w_k^+\right]}j_k,\nonumber\\
\G_{5p}=&\frac{1}{2}g^2C_F\int
\frac{\dk{\vec{k}}\tilde{F}(\vec{p}-\vec{k})}{\w_k^+\w_k^-}
\frac{\s{\vec{p}}{\vec{k}}}{\vec{p\,}^2}
\frac{\left[\w_k^+-\w_k^-\right]}
{\left[M_k^+\w_k^-+M_k^-\w_k^+\right]}g_k,\nonumber\\
\G_{6p}=&\frac{1}{2}g^2C_F\int
\frac{\dk{\vec{k}}\tilde{F}(\vec{p}-\vec{k})}{\w_k^+\w_k^-}
\frac{1}{\vec{p\,}^2}\left(1-\frac{3\Delta}{2}\right)
\left\{
\left[\w_k^++\w_k^-\right]h_k-\left[M_k^++M_k^-\right]j_k
\right\},\nonumber\\
\G_{7p}=&\frac{1}{2}g^2C_F\int
\frac{\dk{\vec{k}}\tilde{F}(\vec{p}-\vec{k})}{\w_k^+\w_k^-}
\frac{1}{\vec{p\,}^2}\left(1-\frac{3\Delta}{2}\right)
\left\{
\frac{\left[M_k^++M_k^-\right]}{\left[\w_k^++\w_k^-\right]}f_k-g_k
\right\},
\end{align}
where we write
\begin{equation}
\Delta=1-\frac{\s{\vec{p}}{\vec{k}}^2}{\vec{p\,}^2\vec{k}^2}.
\label{eq:delta}
\end{equation}
Assuming that the functions $f_k$, $h_k$, $g_k$ and $j_k$ are finite 
(as we shall shortly demonstrate), the vector meson \BS vertex $\G_V$ 
(and its components, $\G_i$) is given as a convolution integral involving 
the IR divergent interaction and the quark mass functions, just as for 
the pseudoscalar meson.  Further, we notice that if we artificially set 
$\Delta=0$, the first four equations are identical to those in the 
pseudoscalar case.  In the special case of equal mass quarks, $\G_2$ and 
$\G_5$ vanish.  Inverting the definitions of $g_k$ and $j_k$, 
\eq{eq:gjdef}, one has
\begin{align}
P_0^2g_p
-\left[\tilde{\w}_p^++\tilde{\w}_p^-\right]\left[\w_p^++\w_p^-\right]j_p
=&(\G_{0p}+\vec{p\,}^2\G_{6p})\left[M_p^+\w_p^-+M_p^-\w_p^+\right]
-\G_{4p}\vec{p\,}^2\left[\w_p^++\w_p^-\right],\nonumber\\
\frac{\left[\tilde{\w}_p^++\tilde{\w}_p^-\right]}
{\left[\w_p^++\w_p^-\right]}g_p-j_p
=&(\G_{1p}
-\vec{p\,}^2\G_{7p})
\frac{\left[M_p^+\w_p^-+M_p^-\w_p^+\right]}{\left[M_p^++M_p^-\right]}
+\G_{5p}\vec{p\,}^2
\frac{\left[\w_p^+-\w_p^-\right]}{\left[M_p^++M_p^-\right]}
\end{align}
in addition to the analogous expressions defined as for the pseudoscalar 
case, \eq{eq:qinv}.  Collecting together the definitions, along with 
\eq{eq:wdiv} for the factors involving $\tilde{\w}$, this results in 
the following four equations:
\begin{align}
h_p=&\frac{P_0^2}{\left[\w_p^++\w_p^-\right]^2}f_p
+\frac{1}{2}g^2C_F\int
\frac{\dk{\vec{k}}\tilde{F}(\vec{p}-\vec{k})}{\w_k^+\w_k^-}
\nonumber\\&\times\left\{
\left[\left(1-\frac{\Delta}{2}\right)
\left[\w_k^++\w_k^-\right]h_k
+\frac{\Delta}{2}\left[M_k^++M_k^-\right]j_k\right]
\frac{\left[M_p^+\w_p^-+M_p^-\w_p^+\right]}
{\left[\w_p^++\w_p^-\right]\left[M_p^++M_p^-\right]}
\right.\nonumber\\&\left.
+h_k\frac{\s{\vec{p}}{\vec{k}}}{\vec{k}^2}
\frac{\left[M_k^+\w_k^--M_k^-\w_k^+\right]\left[M_p^+-M_p^-\right]}
{\left[\w_p^++\w_p^-\right]^2}
-h_p\frac{\s{\vec{p}}{\vec{k}}}{\vec{p\,}^2}
\frac{\left[\w_p^+\w_k^-+\w_p^-\w_k^+\right]}
{\left[\w_p^++\w_p^-\right]}
\right\},\nonumber\\
f_p=&h_p
+\frac{1}{2}g^2C_F\int
\frac{\dk{\vec{k}}\tilde{F}(\vec{p}-\vec{k})}{\w_k^+\w_k^-}\left\{
\left[\left(1-\frac{\Delta}{2}\right)
\frac{\left[M_k^++M_k^-\right]}{\left[\w_k^++\w_k^-\right]}f_k
+\frac{\Delta}{2}g_k\right]
\frac{\left[M_p^+\w_p^-+M_p^-\w_p^+\right]}{\left[\w_p^++\w_p^-\right]}
\right.\nonumber\\&\left.
+f_k\frac{\s{\vec{p}}{\vec{k}}
\left[M_k^++M_k^-\right]}{\left[M_k^+\w_k^-+M_k^-\w_k^+\right]}
-f_p\frac{\s{\vec{p}}{\vec{k}}}{\vec{p\,}^2}
\frac{\left[\w_p^+\w_k^-+\w_p^-\w_k^+\right]}
{\left[\w_p^++\w_p^-\right]}
\right\},\nonumber\\
j_p=&\frac{P_0^2}{\left[\w_p^++\w_p^-\right]^2}g_p
+\frac{1}{2}g^2C_F\int
\frac{\dk{\vec{k}}\tilde{F}(\vec{p}-\vec{k})}{\w_k^+\w_k^-}
\nonumber\\&\times\left\{
\left[\frac{\s{\vec{p}}{\vec{k}}^2}{\vec{p\,}^2\vec{k}^2}
\left[M_k^++M_k^-\right]j_k+\Delta\left[\w_k^++\w_k^-\right]h_k\right]
\frac{\left[M_p^+\w_p^-+M_p^-\w_p^+\right]}{\left[\w_p^++\w_p^-\right]^2}
\right.\nonumber\\&\left.
+j_k\frac{\s{\vec{p}}{\vec{k}}
\left[\w_k^++\w_k^-\right]\left[M_k^++M_k^-\right]}
{\left[\w_p^++\w_p^-\right]\left[M_k^+\w_k^-+M_k^-\w_k^+\right]}
-j_p\frac{\s{\vec{p}}{\vec{k}}}{\vec{p\,}^2}
\frac{\left[\w_p^+\w_k^-+\w_p^-\w_k^+\right]}
{\left[\w_p^++\w_p^-\right]}
\right\},\nonumber\\
g_p=&j_p
+\frac{1}{2}g^2C_F\int
\frac{\dk{\vec{k}}\tilde{F}(\vec{p}-\vec{k})}{\w_k^+\w_k^-}\left\{
\left[\frac{\s{\vec{p}}{\vec{k}}^2}{\vec{p\,}^2\vec{k}^2}g_k
+\Delta\frac{\left[M_k^++M_k^-\right]}
{\left[\w_k^++\w_k^-\right]}f_k\right]
\frac{\left[M_p^+\w_p^-+M_p^-\w_p^+\right]}{\left[M_p^++M_p^-\right]}
\right.\nonumber\\&\left.
+g_k\frac{\s{\vec{p}}{\vec{k}}
\left[\w_k^+-\w_k^-\right]
\left[\w_p^+-\w_p^-\right]}
{\left[M_k^+\w_k^-+M_k^-\w_k^+\right]\left[M_p^++M_p^-\right]}
-g_p\frac{\s{\vec{p}}{\vec{k}}}{\vec{p\,}^2}
\frac{\left[\w_p^+\w_k^-+\w_p^-\w_k^+\right]}
{\left[\w_p^++\w_p^-\right]}
\right\}.
\label{eq:vbses}
\end{align}
As before for the pseudoscalar mesons, the singularities arising from 
$\tilde{F}$ when $\vec{p}=\vec{k}$ cancel, showing that all four 
functions are indeed finite.  In the equal mass case, the equations 
reduce to
\begin{align}
h_p=&\frac{P_0^2}{4\w_p^2}f_p+\frac{1}{2}g^2C_F\int
\frac{\dk{\vec{k}}\tilde{F}(\vec{p}-\vec{k})}{\w_k}\left\{
\left(1-\frac{\Delta}{2}\right)h_k
-h_p\frac{\s{\vec{p}}{\vec{k}}}{\vec{p\,}^2}
+\frac{\Delta}{2}\frac{M_k}{\w_k}j_k
\right\},\nonumber\\
f_p=&h_p+\frac{1}{2}g^2C_F\int
\frac{\dk{\vec{k}}\tilde{F}(\vec{p}-\vec{k})}{\w_k}\left\{
\left(1-\frac{\Delta}{2}\right)\frac{M_pM_k}{\w_k^2}f_k
+\frac{\s{\vec{p}}{\vec{k}}}{\w_k^2}f_k
-f_p\frac{\s{\vec{p}}{\vec{k}}}{\vec{p\,}^2}
+\frac{\Delta}{2}\frac{M_p}{\w_k}g_k
\right\},\nonumber\\
j_p=&\frac{P_0^2}{4\w_p^2}g_p+\frac{1}{2}g^2C_F\int
\frac{\dk{\vec{k}}\tilde{F}(\vec{p}-\vec{k})}{\w_k}\left\{
\frac{\s{\vec{p}}{\vec{k}}^2}{\vec{p\,}^2\vec{k}^2}
\frac{M_pM_k}{\w_p\w_k}j_k
+\frac{\s{\vec{p}}{\vec{k}}}{\w_p\w_k}j_k
-j_p\frac{\s{\vec{p}}{\vec{k}}}{\vec{p\,}^2}
+\Delta\frac{M_p}{\w_p}h_k
\right\},\nonumber\\
g_p=&j_p+\frac{1}{2}g^2C_F\int
\frac{\dk{\vec{k}}\tilde{F}(\vec{p}-\vec{k})}{\w_k}\left\{
\frac{\s{\vec{p}}{\vec{k}}^2}{\vec{p\,}^2\vec{k}^2}
\frac{\w_p}{\w_k}g_k
-g_p\frac{\s{\vec{p}}{\vec{k}}}{\vec{p\,}^2}
+\Delta\frac{\w_pM_k}{\w_k^2}f_k
\right\}.
\label{eq:vbsep}
\end{align}
The above sets of coupled equations have a rather illuminating 
structure.  One sees that outside the integrals, the functions 
($f,h$) and ($g,j$) appear pairwise, each with only one occurrence 
of $P_0^2=M_V^2$.  Inside the integral, the pairing is ($f,g$) and 
($h,j$).  This pairwise connection will be visible in the numerical 
solutions.  Within the integrals, the pairing comes with an associated 
factor of $\Delta$ that serves to couple the two sets of equations for 
the pairings ($f,h$) and ($g,j$) (hence the decoupling when $\Delta$ is 
artificially set to zero).  $\Delta$, as defined in \eq{eq:delta}, is 
dependent only on the angle between the spatial momenta $\vec{p}$ and 
$\vec{k}$.  That when an angular dependence is suppressed, the equations 
for the vector meson reduce to those of the pseudoscalar has an obvious 
interpretation: the emergence of the quantum mechanical description of 
the vector meson as being a total angular momentum excitation of the 
groundstate (pseudoscalar) meson, even in the context of the relativistic 
(though noncovariant) \BS framework used here.

The normalization of the vector meson \BS vertex function can now be 
discussed.  Taking \eq{eq:vnorm}, inserting the vector meson vertex 
decomposition, \eq{eq:vdec}, its conjugate, \eq{eq:cvdec}, along with 
\eq{eq:qdec} for the quark propagator and treating as for the 
pseudoscalar case, the result is
\begin{equation}
1=-\frac{4N_c}{3}\int
\frac{\dk{\vec{k}}}{\w_k^+\w_k^-}
\left[2f_k^Nh_k^N+g_k^Nj_k^N\right]
\frac{\left[M_k^++M_k^-\right]}{\left[M_k^+\w_k^-+M_k^-\w_k^+\right]},
\label{eq:vnormp}
\end{equation}
or, for equal mass quarks
\begin{equation}
1=-\frac{4N_c}{3}\int\frac{\dk{\vec{k}}}{\w_k^3}
\left[2f_k^Nh_k^N+g_k^Nj_k^N\right],
\end{equation}
and where again, only the finite combinations $f_k$, $h_k$, $g_k$ and 
$j_k$ defined in Eqs.~(\ref{eq:fhdef},\ref{eq:gjdef}) appear.  Just as 
for the leptonic decay constants and the \BS equations, the vector meson 
normalization has the same structure as for the pseudoscalar case, 
\eq{eq:norm}.

Finally, let us discuss the heavy quark limit for the vector meson.  This 
proceeds as for the pseudoscalar case.  With the normalization convention 
that
\begin{equation}
g_p^N=\frac{g_p}{\imath{\cal N}},\;\;\;\;j_p^N=\frac{j_p}{\imath{\cal N}}
\end{equation}
($h_p$ and $f_p$ as in \eq{eq:normn} and with 
$h(\vec{p}^2\rightarrow0)=1$), we have from the normalization that 
($m^+=m_h\rightarrow\infty$ as before)
\begin{equation}
{\cal N}^2\stackrel{m^+=m_h}{\longrightarrow}
\frac{4N_c}{3m_h}
\int\frac{\dk{\vec{k}}}{\w_k^-\left[\w_k^-+M_k^-\right]}
\left[2f_kh_k+g_kj_k\right],
\end{equation}
or ${\cal N}\sim1/\sqrt{m_h}$.  The vector meson leptonic decay 
constant, \eq{eq:fv}, becomes
\begin{equation}
f_V\stackrel{m^+=m_h}{\longrightarrow}
\frac{2N_c}{3{\cal N}M_V}\int\frac{\dk{\vec{k}}}{\w_k^-}
\left(2h_k+j_k\right).
\end{equation}
The vector meson \BS equation, \eq{eq:vbses}, gives us the further 
information that $f_p=h_p$, $j_p=g_p$ and $M_V=m_h$ in the heavy quark 
limit, such that as $m_h\rightarrow\infty$,
\begin{equation}
f_V\sqrt{M_V}\sim\mbox{const.}
\label{eq:hlv}
\end{equation}
Again, this agrees with the known heavy quark limit at leading order 
\cite{Neubert:1993mb}.  There is one further result that we should check 
and this concerns the ratio $f_V/f_{PS}$.  The heavy quark effective 
theory result at leading order is \cite{Neubert:1993mb}
\begin{equation}
\frac{f_V}{f_{PS}}=1-\frac{2\al_s(m_h)}{3\pi}
\end{equation}
($\al_s$ being evaluated at the scale $m_h$).  In our case, we do not 
include $\al_s$ contributions, so we should see that for heavy-light 
systems as $m^+=m_h\rightarrow\infty$
\begin{equation}
\frac{f_V}{f_{PS}}\stackrel{m^+=m_h}{\longrightarrow}1.
\label{eq:hlff}
\end{equation}

\section{Numerical implementation}
Before presenting the results, it is worthwhile discussing the numerical 
implementation of the equations.  While the infrared singularities of the 
interaction do cancel formally, they present a formidable challenge for 
the numerical analysis.  The techniques used are based on those employed 
to solve the gap equation for the quark mass function $M_k$, \eq{eq:qgap}, 
explained in detail in Ref.~\cite{Watson:2011kv}.

As a concrete example, let us consider the equal quark mass pseudoscalar 
meson \BS equation for $h_p$, \eq{eq:psbsee}, which reads
\begin{equation}
h_p=\frac{P_0^2}{4\w_p^2}f_p
+\frac{1}{2}g^2C_F\int\frac{\dk{\vec{k}}\tilde{F}(\vec{p}-\vec{k})}{\w_k}
\left\{h_k-h_p\frac{\s{\vec{p}}{\vec{k}}}{\vec{p\,}^2}\right\}
\end{equation}
(all the other \BS equations follow the same pattern).  In conjunction 
with the partner equation for $f_p$, the above is one of a homogeneous 
set of equations, valid only at a particular value of $P_0^2=M_{PS}^2$.  
The standard technique would be to first perform the angular integral 
involving $\tilde{F}(\vec{p}-\vec{k})$ (in this case, the integral can 
be performed analytically).  Then, the radial integral would be set up on 
some discrete grid of $\vec{k}^2$ and $\vec{p\,}^2$ values which results 
in a matrix form for the coupled equations.  One can then use standard 
matrix methods to determine the eigenvalues (the groundstate meson mass 
corresponds to the value of $P_0^2$ where the highest eigenvalue is equal 
to one) and eigenfunctions.  However, in the above case the matrix would 
involve cancellations of singularities ($\tilde{F}$ is divergent when 
$\vec{k}=\vec{p}\,$) between the various terms.  To circumvent this would 
in principle require derivative information about the functions or an 
infrared regularization of $\tilde{F}$ such as a fictitious mass term.  
Further, when $\vec{k}$ is in the vicinity of $\vec{p}$, the matrix 
elements become significantly enhanced.  Keeping control over the 
numerical fidelity in such a situation would become extremely difficult.

As pointed out earlier, the above equation in the chiral limit is very 
similar to that of the quark gap equation, \eq{eq:qgap}, whose numerical 
evaluation is well understood \cite{Watson:2011kv}.  The first step is 
to change integration variables such that the radial momentum flows 
through $\tilde{F}$.  Denoting $\vec{q}=\vec{p}-\vec{k}$, this then gives
\begin{equation}
h_p=\frac{P_0^2}{4\w_p^2}f_p
+\frac{1}{2}g^2C_F\int\frac{\dk{\vec{k}}\tilde{F}_k}{\w_q}
\left\{h_q-h_p\frac{\s{\vec{p}}{\vec{q}}}{\vec{p\,}^2}\right\}.
\end{equation}
Because of the infrared singularity, it is useful to rewrite this 
equation in the form
\begin{equation}
h_p=\frac{X_p+I_p^1}{1+I_p^2},
\end{equation}
where
\begin{equation}
X_p=\frac{P_0^2}{4\w_p^2}f_p,\;\;
I_p^1=\frac{1}{2}g^2C_F\int\frac{\dk{\vec{k}}\tilde{F}_k}{\w_q}h_q,\;\;
I_p^2=\frac{1}{2}g^2C_F\int
\frac{\dk{\vec{k}}\tilde{F}_k}{\w_q}
\frac{\s{\vec{p}}{\vec{q}}}{\vec{p\,}^2}.
\end{equation}
This manipulation substitutes the difference of two infrared singular 
integrals for their ratio, which is numerically less prone to 
instabilities.  Along with the corresponding equation for $f_p$ 
(manipulated in the same way), we now introduce a fictitious eigenvalue 
$\la(P_0^2)$ to the left hand sides of both equations, such that we have 
the form (similarly for $f_p$)
\begin{equation}
\la(P_0^2)h_p=\frac{X_p+I_p^1}{1+I_p^2}.
\end{equation}
Taking a starting guess for the functions $h_p$ and $f_p$ with the 
boundary condition $h(\vec{p\,}^2\rightarrow0)=1$, we construct the 
right hand sides of the equations.  This is doing by interpolating or 
extrapolating $h_q$ and $f_q$ within the angular integrals as required, 
using standard techniques.  As will be seen, the functions $h_p$ and 
$f_p$ are smooth such that this is a numerically safe procedure.  The 
radial integral is performed over a discrete grid of values for 
$\vec{k}^2$, regulated using an infrared cutoff $\e$ and an ultraviolet 
cutoff $\La$ (both with dimensions of $[\mbox{mass}]^2$).  The derived 
value for $h_p$ at the lowest $\vec{p\,}^2$ is now compared with the 
boundary condition (unity) to give the eigenvalue $\la$.  Both derived 
functions $h_p$ and $f_p$ are then scaled by this value such that the 
boundary condition is fulfilled.  The coupled set of equations is then 
iterated until convergence of both the functions and the eigenvalue is 
achieved.  Essentially, this procedure is identical to solving an 
inhomogeneous equation iteratively, but with the lowest grid point value 
for the function (fixed to unity here) swapped for the eigenvalue 
(which for the inhomogeneous equation would be trivially absent).  The 
value of $P_0^2$ for which $\la=1$ is then found.

It is worth pointing out that the numerical procedure described above is 
certainly not without its limitations.  Both integrals $I_p^1$ and 
$I_p^2$ diverge as $1/\sqrt{\e}$ as $\e\rightarrow0$, such that the 
iteration steps are damped (typically by a factor of $10^3$) and one 
must use many steps to achieve a particular tolerance.  However, this 
convergence is stable and easy to implement.  If one were to use the 
difference of the two integrals rather than their ratio, errors in the 
unknown functions are amplified by such factors and the convergence is 
highly unstable -- one would require an extremely precise starting guess 
for the functions in order for the iteration to work.

Once the functions $h_p$ and $f_p$ (and in the case of the vector, $j_p$ 
and $g_p$ too) are found, it is straightforward to obtain the 
normalization and the leptonic decay constant.  In the case of the 
pseudoscalar meson, $h_{PS}$ can also be found in order to verify the 
relation \eq{eq:fhcomp1}.

\section{Results}
Aside from the quark masses, there is in principle only one free 
parameter in the truncation scheme studied in this work and this is the 
string tension $\si$ used in the input function $\tilde{F}$.  By setting 
$\si=1$, all quantities are then (and unless otherwise stated) expressed 
in appropriate units of $\si$.  To compare with physical quantities, we 
also give an estimate of the magnitude using the range 
$\sqrt{\si}\sim\sqrt{\si_W}-\sqrt{3\si_W}\sim440-762\mbox{MeV}$ as 
discussed in Sec.~II.  We use the following numerical parameters 
throughout this section: the infrared cutoff, $\e=10^{-6}$, and the 
ultraviolet cutoff, $\La=10^6$.  Concerning the value of $\e$: in 
Ref.~\cite{Watson:2011kv} it is seen that the quark mass function 
converges for $\e\leq10^{-6}$.  We have checked the results here against 
$\e=10^{-5}$ (given the infrared singularities involved, lower values of 
$\e$ result in significantly more computational effort), and the 
difference is typically $\leq10^{-3}$ ($\leq3\times10^{-3}$) in the 
pseudoscalar (vector) meson masses and $\leq\sim10^{-4}$ 
($\leq3\times10^{-3}$) in the pseudoscalar (vector) meson leptonic decay 
constants.  With the typical range of $\si$ values, this tolerance would 
correspond to $\sim1\mbox{MeV}$.  However, notice that for the quark 
mass function, the difference between results is much smaller for 
$\e=10^{-6}-10^{-7}$ than between $\e=10^{-5}-10^{-6}$ 
\cite{Watson:2011kv}.  Thus, we consider the results presented here to 
be convergent within reasonable limits.  The value of $\La$ is chosen 
such that the heavy quark masses (up to $m=10^3$) can be accommodated: 
with the interaction \eq{eq:int}, all results are UV-finite.

Let us begin with the case of (equal mass) chiral quarks.  Solving 
\eq{eq:qgap} for the quark mass function 
(as in Ref.~\cite{Watson:2011kv}) and inserting into \eq{eq:cond}, the 
chiral condensate is found to be $-0.0123$, which with 
$\sqrt{\si}=440-762\mbox{MeV}$ gives 
$\ev{\ov{q}q}=-(102-176\mbox{MeV})^3$.  While the quark condensate is 
not a physical quantity, this value is considerably lower than the 
commonly accepted $\ev{\ov{q}q}\sim-(250\mbox{MeV})^3$.  The result 
here is consistent with previous similar studies, e.g., 
Refs.~\cite{Adler:1984ri,Pak:2011wu}.  Turning to the pseudoscalar meson, 
this is a special case in that we know analytically that $M_{PS}=0$.  The 
pseudoscalar leptonic decay constant is found to be $f_{PS}=0.0260$, 
which corresponds to $f_{PS}\approx11-20\mbox{MeV}$ with the above range 
for $\sqrt{\si}$.  Again this result agrees with previous studies, e.g., 
Ref.~\cite{Adler:1984ri}.  Comparing with the physical result for the 
pion (with our conventions for the normalization), 
$f_{\pi}=92.2\mbox{MeV}$ \cite{Rosner:2012np}, we see that the 
pseudoscalar leptonic decay constant for light quarks is almost an order 
of magnitude too small.  Clearly, while chiral symmetry is dynamically 
broken by the infrared enhanced interaction, the absence of the spatial 
components of the interaction makes a big difference to the amount of 
symmetry breaking.  The inclusion of more detailed interactions has been 
recently studied within this context in Ref.~\cite{Pak:2011wu}.

Turning to the chiral quark vector meson, we find that $M_{V}=1.553$, or 
$M_V=683-1184\mbox{MeV}$.  This has the correct magnitude when compared 
to the physical result, $M_{\ro}=775.5\mbox{MeV}$ 
\cite{Beringer:1900zz}.  The leptonic decay constant is found to be 
$f_V=0.250$, or $f_V=110-190\mbox{MeV}$, as compared to 
$f_\ro=153\mbox{MeV}$ \cite{Maris:1999nt} 
(with our normalization convention).  The calculated vector meson 
leptonic decay constant is thus also comparable with the magnitude of 
the physical result, unlike its pseudoscalar counterpart.

\begin{figure}[t]
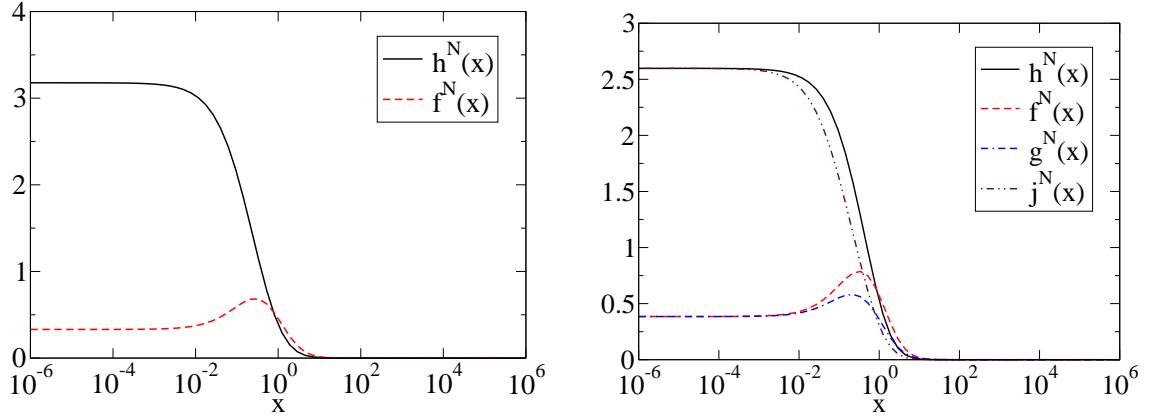

\vspace{0.8cm}
\includegraphics[width=0.4\linewidth]{reseqvertps.eps}
\hspace{0.5cm}
\includegraphics[width=0.4\linewidth]{reseqvertv.eps}
\vspace{0.3cm}
\caption{\label{fig:eqvert}[left panel] Pseudoscalar and [right panel] 
vector normalized vertex functions with (equal) chiral quarks, plotted 
as a function of $x=\vec{k}^2$.  All dimensionfull quantities are in 
appropriate units of the string tension, $\si$.  See text for details.}
\end{figure}

Staying with the chiral mesons, the normalized vertex functions $h^N$, 
$f^N$ and for the vector, $g^N$ and $j^N$, are plotted in 
Fig.~\ref{fig:eqvert}.  Recall that these functions are constructed so 
as to be free from singularities and this is explicitly verified 
(and has been checked for all combinations of quark masses).  The 
functions are smooth and well-behaved such that the numerical 
implementation of the angular integrals is indeed justified 
(see the previous section).  In the chiral limit, where $M_{PS}=0$, 
the pseudoscalar function $h^N$ should be identical to the chiral quark 
mass function, $M$, up to a constant factor and this has been verified 
(and serves as a useful numerical check).  The vector functions $h^N$ 
and $j^N$ have a similar form, but with minor differences.  The 
functions $f^N$ (pseudoscalar and vector) and $g^N$ exhibit a bump at 
around $x=\vec{k}^2\sim0.2-0.5$, although this does not seem to have 
any significance.  In the vector case, the pairwise behavior of the 
functions ($f$, $g$) and ($h$, $j$) is clearly visible in the IR.  As 
mentioned in the discussion following \eq{eq:vbsep}, these pairings 
occur in the integral expressions for the vector meson \BS equation: 
the pairs have the same infrared limit, but at finite momenta $j^N$ 
and $g^N$ are generally smaller and shifted slightly towards infrared 
momenta.

\begin{figure}[t]
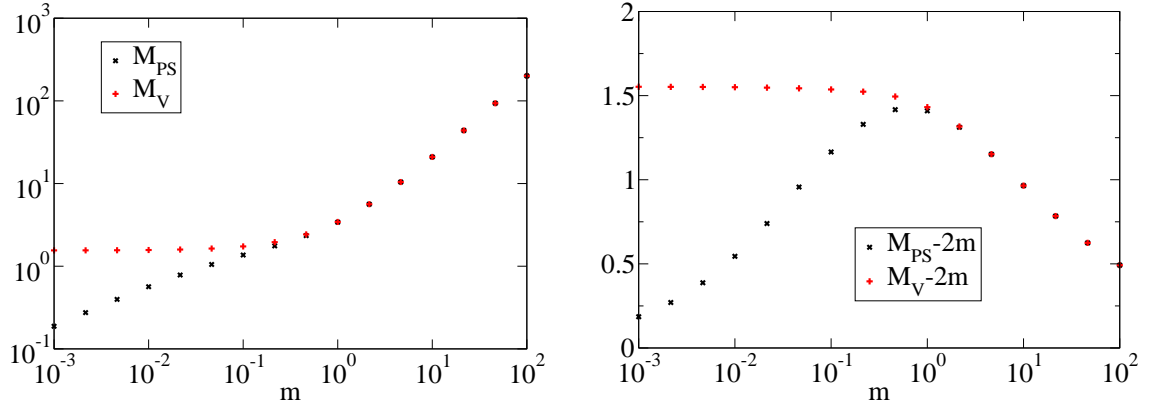

\vspace{0.8cm}
\includegraphics[width=0.4\linewidth]{reseqml.eps}
\hspace{0.5cm}
\includegraphics[width=0.4\linewidth]{reseqe.eps}
\vspace{0.3cm}
\caption{\label{fig:eqm}Pseudoscalar and vector meson masses 
[left panel] and binding energies [right panel] with equal mass quarks, 
plotted as a function of the quark mass.  All dimensionfull quantities 
are in appropriate units of the string tension, $\si$.  See text for 
details.}
\end{figure}

\begin{figure}[t]
\vspace{0.8cm}
\includegraphics[width=0.5\linewidth]{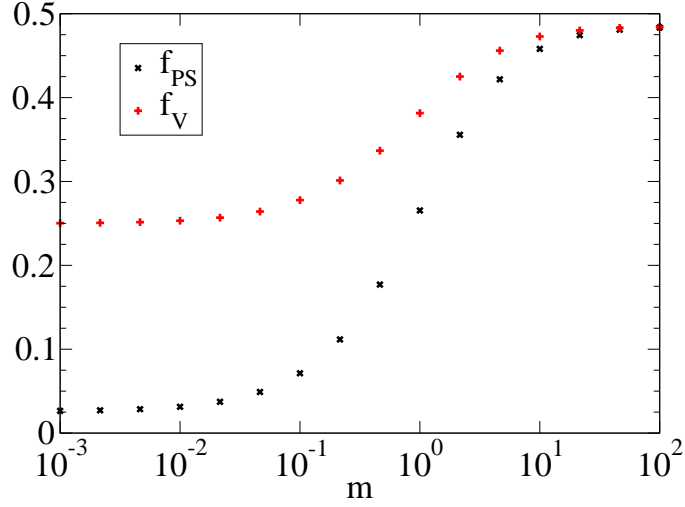}
\vspace{0.3cm}
\caption{\label{fig:eqf}Pseudoscalar and vector meson leptonic decay 
constants with equal mass quarks, plotted as a function of the quark 
mass.  All dimensionfull quantities are in appropriate units of the 
string tension, $\si$.  See text for details.}
\end{figure}

Let us now consider mesons with equal mass quarks.  The meson masses are 
plotted in Fig.~\ref{fig:eqm} as a function of the quark mass, $m$.  One 
can see that the pseudoscalar meson mass goes approximately like 
$\sqrt{m}$ in the chiral limit, a clear signal for the dynamical 
breaking of chiral symmetry and the Goldstone boson nature of the pion.  
For large quark masses, both meson masses approximate to $2m$ and are 
degenerate.  We have verified that \eq{eq:fhcomp1} is numerically 
satisfied (this again serves as a useful numerical check).  These 
results are in explicit agreement with those presented in 
Ref.~\cite{Alkofer:2005ug}.  The binding energy, defined as $M_{PS}-2m$ 
(similarly for the vector meson), is also plotted.  This provides a more 
detailed picture of the interaction, since the meson mass is dominated 
by large quark masses in that limit.  One can see that the binding energy 
is not trivial and decreases for larger quark masses.  The degeneracy of 
the meson masses for heavy quarks is also clearly visible from the 
binding energy.  That the pseudoscalar and vector meson masses are 
almost equal for equal quark masses above ${\cal O}(1)$, means that the 
mass splitting of charmonium and bottomonium is not present.  The 
leptonic decay constants for the equal quark mass mesons are plotted in 
Fig.~\ref{fig:eqf}.  Unlike the meson masses, the leptonic decay 
constants are only degenerate in the heavy quark limit.  For 
bottomonium, recent lattice results give $f_{PS}=472\mbox{MeV}$ 
\cite{McNeile:2012qf} (with our conventions); comparing with our 
results, we get (for the heaviest quark mass) at most $f_{PS}\sim0.48$, 
or $f_{PS}\sim211-366\mbox{MeV}$.  The vector meson result is similar, 
meaning that both leptonic decay constants are significantly too small, 
even taking into account the range of values for $\si$.  In the equal 
quark mass case for large quark masses, the pseudoscalar and vector 
normalizations become degenerate and go to zero; further, all vertex 
functions become degenerate.

\begin{figure}[t]
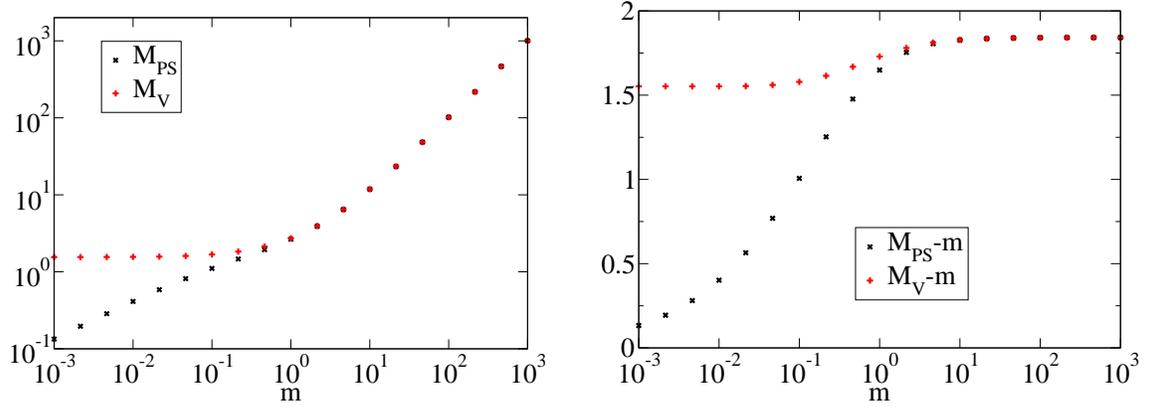

\vspace{0.8cm}
\includegraphics[width=0.4\linewidth]{reshlm.eps}
\hspace{0.5cm}
\includegraphics[width=0.4\linewidth]{reshle.eps}
\vspace{0.3cm}
\caption{\label{fig:hlm}Pseudoscalar and vector meson masses 
[left panel] and binding energies [right panel] with one fixed chiral 
quark, plotted as a function of the other quark mass.  All dimensionfull 
quantities are in appropriate units of the string tension, $\si$.  
See text for details.}
\end{figure}

\begin{figure}[t]
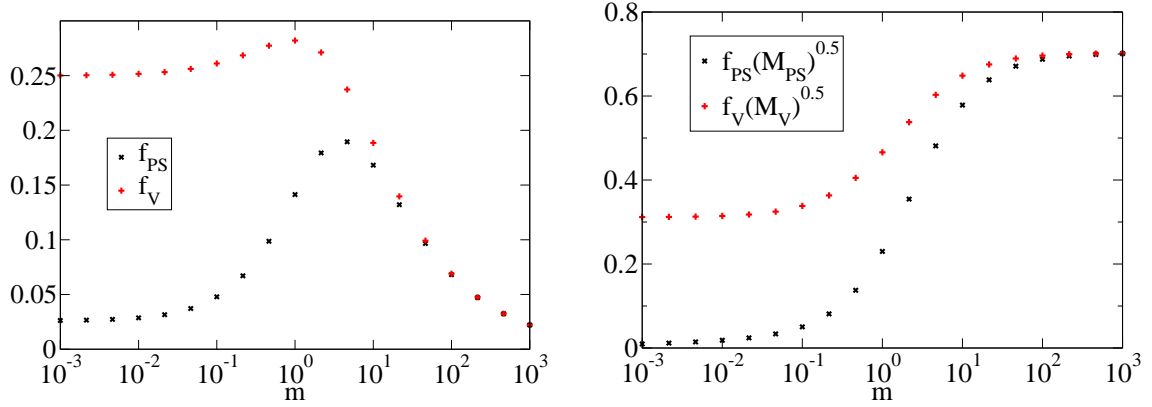

\vspace{0.8cm}
\includegraphics[width=0.4\linewidth]{reshlf.eps}
\hspace{0.5cm}
\includegraphics[width=0.4\linewidth]{reshlfm.eps}
\vspace{0.3cm}
\caption{\label{fig:hlf}Pseudoscalar and vector meson leptonic decay 
constants [left panel] and $f_{PS}\sqrt{M_{PS}}$, $f_V\sqrt{M_V}$ 
[right panel] with one fixed chiral quark, plotted as a function of the 
other quark mass.  All dimensionfull quantities are in appropriate units 
of the string tension, $\si$.  See text for details.}
\end{figure}

\begin{figure}[t]
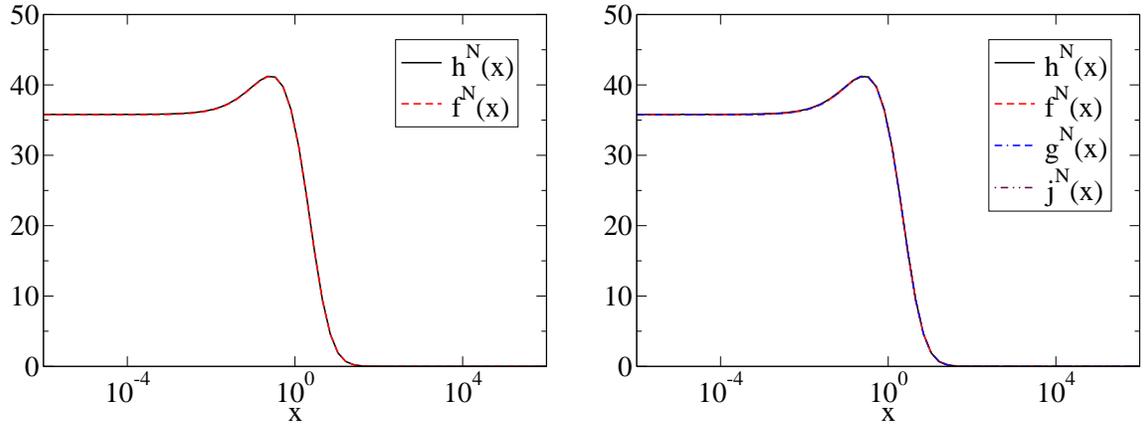

\vspace{0.8cm}
\includegraphics[width=0.4\linewidth]{reshlvertps.eps}
\hspace{0.5cm}
\includegraphics[width=0.4\linewidth]{reshlvertv.eps}
\vspace{0.3cm}
\caption{\label{fig:hlvert}[left panel] Pseudoscalar and 
[right panel] vector normalized vertex functions with one chiral quark 
and one quark of mass $10^3$, plotted as a function of $x=\vec{k}^2$.  
All dimensionfull quantities are in appropriate units of the string 
tension, $\si$.  See text for details.}
\end{figure}

Finally, let us consider the case where one quark has fixed mass.  The 
meson masses are plotted in Fig.~\ref{fig:hlm} for the case where one 
quark is fixed to zero bare mass (i.e., $m^-=0$, $m^+=m$).  In the limit 
as the other quark mass ($m$) vanishes, $M_{PS}\sim\sqrt{m}$ 
(signaling dynamical chiral symmetry breaking), whereas for large 
masses, $M_{PS}\approx M_V\sim m$ as one would expect.  Again, 
\eq{eq:fhcomp1} has been numerically verified.  As previously, defining 
a binding energy as $M_{PS}-m$ (similarly for the vector) reveals the 
interaction in slightly more detail and this is also shown in 
Fig.~\ref{fig:hlm}.  The meson masses are degenerate above 
$m\sim{\cal O}(10)$ and the mass-splitting between states is again too 
small in this limit.  Unlike the case of equal mass quarks we see that 
the binding energy does not decrease for large quark masses.  Turning to 
the leptonic decay constants, plotted in Fig.~\ref{fig:hlf}, the results 
are again consistently too small.  For example, $f_{B}=144\mbox{MeV}$ 
\cite{Hwang:2010hw,Rosner:2008yu} (with our convention).  Using the 
input value $\sqrt{\si}=\sqrt{\si_W}$ and a quark mass of $10$, 
corresponding to a bottom quark of $m=4.4GeV$ in these units, one has 
$f_{PS}=0.168\sim74\mbox{MeV}$, which is clearly too low.  On the other 
hand, taking $\sqrt{\si}=\sqrt{3\si_W}$ and a quark mass of $4.64$, 
corresponding to $m=3.5\mbox{GeV}$, one has the result 
$f_{PS}=0.190\sim145\mbox{MeV}$: at first glance this looks like a nice 
result, but given that the bottom quark mass (in physical units) is 
realistically too low and that the leptonic decay constant decreases 
from this point onwards as the quark mass increases 
(see Fig.~\ref{fig:hlf}), the result would be also too small.  However, 
the real interest lies in the heavy quark asymptotic behavior of the 
leptonic decay constants.  In both cases, we should see that the 
leptonic decay constant decreases as one over the square root of the 
meson mass and this is explicitly verified in the right panel of 
Fig.~\ref{fig:hlf}, where it is seen that 
Eqs.~(\ref{eq:hlps},\ref{eq:hlv}) are fulfilled.  Further, it is clear 
that \eq{eq:hlff} is satisfied.  Looking at the vertex functions in the 
most extreme case ($m=10^3$), plotted in Fig.~\ref{fig:hlvert}, one sees 
that the functions are all identical.  The heavy quark analysis told us 
that $f^N=h^N$ and $j^N=g^N$; seemingly the pairing of $h^N$ with $j^N$ 
and $f^N$ with $g^N$ for the vector meson, seen for the equal chiral 
quark case, also happens here.  Thus, the numerical solution to the 
truncated system does indeed reflect the heavy quark symmetry.

\section{Summary and conclusions}
In this study, a leading order truncation scheme developed in the 
context of Coulomb gauge \DS equations \cite{Watson:2011kv} has been 
extended to the \BS equation framework for quark-antiquark pseudoscalar 
and vector meson masses and leptonic decay constants, with quark content 
of arbitrary mass.  The leading order Coulomb gauge truncation scheme 
centers around the inclusion of an instantaneous and IR singular 
interaction corresponding to a pure linear rising potential, 
supplemented by a term arising from the  consideration of total color 
charge conservation.  As was briefly discussed, this results in a 
particular form for the quark gap equation 
(the Adler-Davis gap equation \cite{Adler:1984ri}) for the static quark 
mass function and for which the divergences cancel and dynamical chiral 
symmetry breaking is observed.  The quark proper two-point function and 
the self-energy are divergent, reflecting the physical picture that one 
would require an infinite amount of energy to create a colored object 
from the vacuum.  Further, under this truncation scheme, a connection 
to the Coulomb gauge rest frame heavy quark limit (in the absence of pure 
Yang-Mills corrections) \cite{Popovici:2010mb} can be established.

The connection of the quark propagator to the quark-antiquark 
(flavor nonsinglet) \BS equation is well understood via the axialvector 
Ward-Takahashi identity (AXWTI).  Following the approach of 
Ref.~\cite{Maris:1997hd} (written within the context of the covariant 
formalism), the necessary framework was briefly reviewed and adapted to 
the (noncovariant) rest frame Coulomb gauge formalism studied here.  The 
importance and usefulness of the AXWTI is primarily in that it 
guarantees that the pion emerges as the massless Goldstone of dynamical 
chiral symmetry breaking in the chiral limit.  Indeed, earlier studies 
in Coulomb gauge were focused on this aspect 
\cite{Govaerts:1983ft,Alkofer:1988tc}.  However, the AXWTI also applies 
to arbitrary quark masses, including heavy quarks and allowing a 
simultaneous analysis of both the chiral and heavy quark limits.

The necessary formalism for calculating the pseudoscalar and vector 
meson masses and leptonic decay constants was derived.  Given the IR 
singular interaction along with the associated divergence arising from 
the consideration of total color charge conservation, the overriding 
concern was the elimination of the IR divergences within the equations.  
It was seen that this can be explicitly achieved for arbitrary quark 
mass content and that the pseudoscalar meson is described, within this 
Coulomb gauge truncation scheme, by two expressly finite functions; the 
vector meson correspondingly described by four finite functions.  In 
contrast, the \BS vertices themselves are manifestly divergent.  This 
situation is directly analogous to the case of the quark with a finite 
mass function though divergent self-energy.  Such cancellations have 
also been observed in the Coulomb gauge rest frame heavy quark limit for 
mesons \cite{Popovici:2010mb}, baryons \cite{Popovici:2010ph}, and in 
the four-point quark Green's functions \cite{Popovici:2011yz}.  The 
connection to the heavy quark limit was presented and it was shown that 
the leading order results of heavy quark effective theory 
\cite{Neubert:1993mb} are reproduced.

The \BS equation reduces to a particularly intriguing form within this 
truncation scheme.  As discussed, the expressions, 
Eqs.~(\ref{eq:psbses},\ref{eq:vbses}), are a generalization of earlier 
Coulomb gauge studies of the pseudoscalar meson 
\cite{Govaerts:1983ft,Alkofer:1988tc} to arbitrary quark masses and 
extended to include vector mesons.  Given the fact that the \BS equation 
is constructed from the quark proper two-point function via the AXWTI, 
it is no surprise that the equations for the pseudoscalar meson have a 
very similar structure to the quark gap equation 
\cite{Adler:1984ri,Watson:2011kv}, and directly exhibits the chiral 
limit properties that one would expect.  The dependence on the bound 
state energy arises in simple fashion and the vector meson appears as 
an angular momentum excitation of the pseudoscalar.  In contrast to 
covariant gauge studies of the \BS equation 
(e.g., Refs.~\cite{Maris:1997tm,Maris:1999nt,Alkofer:2002bp}) there are 
no complications, at least within this leading order truncation scheme, 
arising from the necessity to evaluate the quark propagator dressing 
functions at complex Euclidean momenta (i.e., the timelike regime) and 
dealing numerically with possible non-analytic structures 
(such as in 
Refs.~\cite{Bhagwat:2002tx,Fischer:2005en,
Krassnigg:2009gd,Fischer:2008sp}).  This allows for an analysis, both 
analytic and numerical, of mesons with arbitrarily asymmetric quark 
content.

The derived \BS equations were solved numerically and results for the 
pseudoscalar and vector meson masses and their respective leptonic decay 
constants were presented for arbitrary quark masses.  It was explicitly 
verified that the IR singularities cancel.  Clearly visible were both the 
qualitative pattern of dynamical chiral symmetry breaking and the leading 
order heavy quark limit.  In this respect, the leading order truncation, 
despite its lack of sophistication is rather successful.  
Quantitatively, the degree of dynamical mass generation, the mass 
splitting between the pseudoscalar and vector meson masses, and the 
leptonic decay constants are all too small (although the vector meson 
mass and the leptonic decay constant in the chiral limit do have 
roughly the correct order of magnitude).  Having focused on the 
cancellation of infrared divergences in this study, the coupling of 
the quarks to the spatial components of the gluon field and the gluon 
self-interaction have been neglected within the context of the leading 
order truncation scheme.  Clearly, a quantitative comparison to the 
physical spectrum should include such interactions and this we propose 
to do in future work.

\begin{acknowledgments}
This work has been supported by the Deutsche Forschungsgemeinschaft 
(DFG) under contracts no. DFG-Re856/6-2,3.
\end{acknowledgments}

\end{document}